\documentclass[twocolumn,numberedappendix,tighten]{aastex6}

\usepackage{amsmath,amssymb,amstext}
\usepackage[all]{hypcap} 
\usepackage{natbib}
\usepackage[multidot]{grffile}
\usepackage{longtable}

\usepackage{graphicx} %
\newcommand{\suz}{{\it Suzaku }}
\newcommand{\xmm}{{\it XMM-Newton }}
\newcommand{\chan}{{\it Chandra }}

\slugcomment{Submitted to ApJ, 2016 April 4}

\begin{document}

\title{Radial Profile of the 3.5 keV line  out to $r_{200}$ in the Perseus Cluster}
\author{Jeroen~Franse\altaffilmark{1,2},
Esra~Bulbul\altaffilmark{3},
Adam~Foster\altaffilmark{4},  
Alexey~Boyarsky\altaffilmark{2},   
Maxim~Markevitch\altaffilmark{5}, 
Mark~Bautz\altaffilmark{3},
Dmytro~Iakubovskyi\altaffilmark{6,7},
Mike~Loewenstein\altaffilmark{8,9},
Michael~McDonald\altaffilmark{3},
Eric~Miller\altaffilmark{3},   
Scott~W.~Randall\altaffilmark{4}, 
Oleg~Ruchayskiy\altaffilmark{6},
Randall~K.~Smith\altaffilmark{4}  
}

\altaffiltext{1}{Leiden Observatory, Leiden University, Niels Bohrweg 2, Leiden, The Netherlands}
\altaffiltext{2}{Instituut-Lorentz for Theoretical Physics, Leiden University, Niels Bohrweg 2, Leiden, The Netherlands}
\altaffiltext{3}{Kavli Institute for Astrophysics \& Space Research, Massachusetts Institute of Technology, 77 Massachusetts Ave., Cambridge, MA 02139, USA}
\altaffiltext{4}{Harvard-Smithsonian Center for Astrophysics, 60 Garden Street, Cambridge, MA~02138, USA}
\altaffiltext{5}{NASA Goddard Space Flight Center, 880 Greenbelt Rd., MD 20771, USA}
\altaffiltext{6}{Discovery Center, Niels Bohr Institute, Blegdamsvej 17, Copenhagen, Denmark}
\altaffiltext{7}{Bogolyubov Institute of Theoretical Physics, Metrologichna, Str. 14-b, 03680, Kyiv, Ukraine}
\altaffiltext{8}{CRESST and X-ray Astrophysics Laboratory NASA/GSFC, Greenbelt, MD 20771}
\altaffiltext{9}{Department of Astronomy, University of Maryland, College Park, MD 20742}

\begin{abstract}
The recent discovery of the unidentified emission line at 3.5~keV in galaxies and clusters has attracted great interest from the community. As the origin of the line remains uncertain, we study the surface brightness distribution of the line in the Perseus cluster since that information can be used to identify its origin. 
We examine the flux distribution of the 3.5 keV line in the deep \suz observations of the Perseus cluster in detail. 
The 3.5~keV line is observed in three concentric annuli in the central observations, although the observations of the outskirts of the cluster did not reveal such a signal.
We establish that these detections and the upper limits from the non-detections are consistent with a dark matter decay origin. 
However, absence of positive detection in the outskirts is also consistent with some unknown astrophysical origin of the line in the dense gas of the Perseus core, as well as with a dark matter origin with a steeper dependence on mass than the dark matter decay. We also comment on several recently published analyses of the 3.5~keV line.\\

\end{abstract}
\maketitle

\section{Introduction}
The recent discovery of the unidentified X-ray line at $\sim$3.5 keV in the stacked {\it XMM-Newton} and {\it Chandra} observations of 73 galaxy clusters and in M31 and its possible interpretation as a decaying dark matter have attracted great attention from the community (\cite{Bulbul:14,Boyarsky:14a}, Bu14 and Bo14 respectively from here on). The signal is significantly detected in the center of Perseus (the X-ray brightest cluster on the sky) by the {\it XMM-Newton} and {\it Chandra} satellites \citep[and later confirmed with {\it Suzaku}; see][]{Urban:14} and in its outskirts with {\it XMM-Newton} (Bo14). The signal is also observed in the Galactic Center \citep{Boyarsky:14b,Jeltema:14}.

Although there has been an extensive effort in the community, the origin of the line is still quite uncertain. Among  the three possible interpretations of the line are an instrumental feature, an astrophysical line (e.g., from the intracluster plasma), and emission from dark matter decay or annihilation processes. 
An instrumental line or calibration errors as possible origins of the 3.5~keV line are extensively studied in the original discovery papers by Bu14 and Bo14. Bu14's analysis, in particular, argues that stacking blue-shifted spectra of a large sample of galaxy clusters with a wide redshift range excludes the instrumental artifact. The detection of the line by several detectors on board of  {\it Chandra}, {\it XMM-Newton}, and {\it Suzaku} indicates that it is unlikely due to an instrumental artifact. Furthermore, non-detections in deep exposures of `blank-sky' background observations with {\it XMM-Newton} and {\it Suzaku} also exclude an instrumental artifact \citep[Bo14;][]{Sekiya:2015jsa}.

Another possible interpretation of the $\sim$3.5 keV line is spectral confusion with one of a number of nearby weak astrophysical lines of K~{\sc xviii}, Cl~{\sc xvii}, and Ar~{\sc xvii}, or possible lines from charge exchange in the intra-cluster medium. This has been extensively discussed in Bu14. 
Atomic transitions, specifically from the K~{\sc xviii} and Ar~{\sc xvii} ions are hard to unambiguously distinguish from the 3.5~keV line due to the instruments' spectral resolution (CCD resolution is 100--120 eV FWHM at this energy). Bu14 report that abundances of a 10--20 times solar are required to explain the 3.5~keV excess with any of these lines based on the estimates obtained from the observed S and Ca line ratios. \citet{Jeltema:14b,Jeltema:14} and \citet{Carlson:14} argue that an atomic transition from K~{\sc xviii} in cool $<$1~keV plasma is likely to be responsible for the 3.5~keV line. In a comment to these studies, \cite{Bulbul:14b} showed that the observed line ratios are inconsistent with the existence of any significant quantities of cool gas in clusters used in the Bu14 sample. We address further issues with the updated paper by \citet{Jeltema:14} and \citet{Carlson:14} in Appendix~\ref{app:carlson}.
A recent study by \citep{gu:15} suggests an alternative explanation for the line, i.e. charge exchange with bare sulfur ions at 3.48 keV. This interpretation is discussed in Appendix~\ref{app:cx}.


A more exotic explanation of the 3.5~keV line is emission from decaying dark matter (Bu14; Bo14; \cite{Boyarsky:14a,Boyarsky:14c}). Although the line intensity in the Perseus cluster core appears to be five times brighter than the flux in the stacked clusters if one scales the predicted fluxes with cluster mass as expected for dark matter decay (see Bu14), the relative intensities between other objects (M31, Galactic Center, clusters), and the surface brightness distribution within the Perseus cluster (from \xmm measurements outside the core) are consistent with a decaying dark matter feature \citep{Boyarsky:14a, Boyarsky:14b}. 
The detection in the Galactic center is consistent with the decaying dark matter interpretation, although this result does not exclude K~{\sc xviii} as a possible origin \citep{Boyarsky:14b}. 
The upper limits derived from the blank-sky observations (since these contain dark matter in the field of view from the Galaxy's dark matter halo) are consistent with the fluxes reported by previous studies.
On the other hand, non-detections in several other studies, for instance, in stacked galaxies \citep{Anderson:14} and in dwarf galaxies \citep{Malyshev:14} challenge the decaying dark matter interpretation of the line. However, the reported statistical tensions across these objects are mild, at a level of 2--3$\sigma$ (with the exception of the stacked galaxies). 
Recently, \cite{Ruchayskiy:2015onc} reported on the analysis of newly obtained very-long-exposure \xmm data of the Draco satellite galaxy. A small hint of $\sim$3.5~keV emission was identified although the authors conservatively focus on the upper limits and determine that it is consistent with a decaying dark matter origin based on the dark matter content of the object.
In another work regarding the same Draco data, \cite{Jeltema:2015mee} claim a much stronger limit on the possible $\sim$3.5~keV line flux that is at odds with a dark matter decay interpretation. \cite{Ruchayskiy:2015onc} suggests mainly that their more thorough spectral modeling provides a more accurate continuum model. Primary differences include additional physically motivated model components and a wider spectral fitting range (Iakubovskyi et al. (in prep.) offers a quantitative description of this effect). This influences the line flux limits and brings them in agreement with the previous detections of the 3.5 keV line. 
Most recently, \cite{Bulbul:2016b} reported a weak spectral excess around 3.5~keV in the stacked {\it Suzaku} observations of 47 galaxy clusters. The upper limits derived from their analysis are consistent with the detection from the stacked clusters. However, their sample excludes the Perseus cluster which is in tension with the previously reported line flux observed with {\it XMM-Newton}.

In this work we take a further step to examine the spatial distribution of the 3.5~keV line within the Perseus cluster from its core to outskirts with {\it Suzaku}. The 3.5~keV line is detected in the observations of the core of the Perseus cluster in both the central 6$^\prime$ and in the surrounding area within {\it Suzaku}'s field-of-view by \cite{Urban:14}. The authors confirm the finding of Bu14 that the flux of the 3.5~keV line in the core is too strong for a decaying dark matter interpretation that assumes a single spherical dark matter distribution for the cluster (as measured by \cite{Simionescu:11}). \cite{Urban:14} also studied 3 other clusters observed with {\it Suzaku}, and did not detect any 3.5~keV line flux in them. These non-detections are consistent with the previous results for other clusters and samples \citep[Bu14; Bo14;][]{Boyarsky:14b}. We note that \cite{Tamura:2014mta} also studied the same \suz observations of Perseus, but do not find evidence of excess emission around 3.5~keV; the origin of this discrepancy is unclear and we will discuss it below. 

We here present the analysis of additional \suz data that extend the previous studies to greater radii. This paper is organized as follows: in Section \ref{sec:reduction}, we describe the {\it Suzaku} data reduction and analysis. In Section \ref{sec:results}, we provide our results in the cluster center and in the outskirts. We discuss systematic errors that are relevant to the {\it Suzaku} X-ray measurements at large radii in Section \ref{sec:systematics}. In Sections \ref{sec:discussion} and \ref{sec:conclusion} we discuss our results and present our conclusions. Throughout the paper, a standard $\Lambda$CDM cosmology with H$_{0}$ = 70 km s$^{-1}$ Mpc$^{-1}$, $\Omega_\Lambda$ = 0.7, and $\Omega_M$ = 0.3 is assumed. In this cosmology, 1$^\prime$ at the distance of the cluster corresponds to $\sim$ 21.2 kpc. Unless otherwise stated, reported errors correspond to 68\% (90\%) confidence intervals.

\section{Data Reduction and Analysis}
\label{sec:reduction}

The Perseus cluster has been observed with {\it Suzaku} between 2006 and 2015 for a total 2.3~Ms. We process the {\it Suzaku} data with HEASOFT version 6.13, and the latest calibration database CALDB as of May 2014. The raw event files are filtered using the FTOOL {\it aepipeline}. The detailed steps of the data processing and filtering are given in \cite{Bulbul:15}. The {\it Suzaku} observations utilized in this work and net exposure times of each pointing after filtering are given in Table \ref{table:obs}. 

Point sources in the FOV are detected from the {\it Suzaku} data using CIAO's {\it wavdetect} tool. The detection is performed using {\it  Suzaku}'s half-power radius of 1$^{\prime}$ as the wavelet radius as described in \citep{Urban:14}. The detected point sources are excluded from further analysis. Spectra are extracted from the filtered event files in {\it XSELECT}. 
Corresponding detector redistribution function (RMF) and ancillary response function (ARF) files are constructed using the {\it xisrmfgen} and {\it xisarfgen} tools. The Night-Earth background spectra are generated using the {\it xisnxbgen} tool and subtracted from each total spectrum prior to fitting. 

We co-add front-illuminated (FI) XIS0 and XIS3 data to simplify spectral fitting using FTOOL {\it mathpha}. The back-illuminated (BI) XIS1 data are co-added separately. The exposure-weighted and normalized ARFs and RMFs are stacked using the FTOOLS {\it addarf} and {\it addrmf}. The NXB subtracted FI and BI observations are modeled simultaneously in the 1.95 to 6 keV energy band. 
Following the same approach of Bu14, we model the FI and BI observations with the line-free multi-temperature {\it apec} models and additional Gaussian models for all the relevant atomic transitions, to allow maximum modeling freedom within physical reason. The free parameters of the model are tied between the FI and BI observations. XSPEC v12.9 is used to perform the spectral fits with the ATOMDB version 2.0.2 \citep{Foster:2012hy}. The galactic column density is frozen at the Leiden/Argentine/Bonn (LAB) Galactic HI Survey \citep{Kalberla:05} value of 1.36$\times10^{20}$  cm$^{-2}$ in our fits. Two wide instrumental Au M edges are modeled with two {\it gabs} components  at 2.3 and 3.08 keV following \citet{Tamura:2014mta}.

{
\begin{table*}[ht!]
\begin{center}
\caption{\suz observations of the Perseus cluster utilized in this study. {\it d} indicates the distance from the cluster center in arcminutes}
\renewcommand{\arraystretch}{1.5}
\begin{tabular}{lccclccclccc}
\hline\hline\\
	 ObsID 	& FI	& BI	& $d$ 	& ObsID 	& FI	& BI	& $d$ 	& ObsID 	& FI	& BI	& $d$ 	\\
			& Exp (ks) & 	Exp (ks) & arcmin && Exp (ks) & 	Exp (ks) & arcmin & & Exp (ks) & 	Exp (ks) & arcmin \\
\\\hline
  101012020 &    79.9   &    39.9 & 0  & 804057010 & 24.1 & 12.0 & 32.80 & 806129010 & 12.9 & 6.4 & 75.36\\ 
  102011010 &    70.2   &    35.1 & 0  & 806136010 & 13.1 & 6.5 & 32.81 & 804067010 & 43.9 & 22.0 & 81.63\\ 
  102012010 &    107.0  &    53.5 & 0  & 805104010 & 13.9 & 6.9 & 32.88 & 806118010 & 27.1 & 13.6 & 81.79\\ 
  103004010 &    68.2   &    34.1 & 0  & 806124010 & 19.1 & 9.5 & 33.11 & 806106010 & 24.7 & 12.4 & 82.67\\ 
  103004020 &    92.6   &    46.3 & 0  & 801049040 & 15.0 & 7.5 & 33.12 & 805100010 & 18.9 & 9.5 & 82.82\\ 
  104018010 &    33.9   &    17.0 & 0  & 801049010 & 50.3 & 25.2 & 35.94 & 805107010 & 15.2 & 7.6 & 83.11\\ 
  104019010 &    67.2   &    33.6 & 0  & 806113010 & 19.1 & 9.5 & 40.25 & 804060010 & 43.2 & 21.7 & 83.15\\ 
  105009010 &    59.2   &    29.6 & 0  & 806101010 & 19.5 & 9.7 & 40.86 & 806142010 & 31.6 & 15.8 & 83.23\\ 
  105009020 &    66.0   &    33.0 & 0  & 806137010 & 21.0 & 10.5 & 41.23 & 806130010 & 27.5 & 13.7 & 83.55\\ 
  106005010 &    68.2   &    34.1 & 0  & 806125010 & 11.1 & 5.6 & 41.72 & 808087010 & 34.8 & 17.4 & 87.97\\ 
  106005020 &    68.5   &    41.1 & 0  & 804065010 & 24.5 & 12.2 & 48.03 & 806119010 & 32.5 & 16.3 & 90.56\\ 
  107005010 &    66.4   &    33.2 & 0  & 806114010 & 16.3 & 8.2 & 48.21 & 805111010 & 13.1 & 6.5 & 91.08\\ 
  107005020 &    60.5   &    35.6 & 0  & 805098010 & 13.5 & 6.7 & 49.02 & 806107010 & 30.4 & 15.2 & 91.42\\ 
  108005010 &    62.5   &    38.1 & 0  & 806102010 & 14.4 & 7.2 & 49.05 & 805115010 & 19.5 & 9.7 & 91.53\\ 
  108005020 &    68.2   &    34.1 & 0  & 804058010 & 22.8 & 11.5 & 49.58 & 806143010 & 19.6 & 9.8 & 91.60\\ 
804063010 & 26.9 & 13.5 & 14.48 & 806138010 & 19.7 & 9.9 & 49.59 & 806131010 & 27.9 & 13.9 & 92.00\\ 
806111010 & 21.6 & 10.8 & 14.70 & 805105010 & 21.8 & 10.9 & 49.61 & 804068010 & 60.2 & 30.1 & 98.38\\ 
805096010 & 16.3 & 8.1 & 15.54 & 806126010 & 15.0 & 7.5 & 49.93 & 806120010 & 17.1 & 8.6 & 98.57\\ 
806099010 & 23.1 & 11.6 & 15.58 & 806115010 & 23.8 & 11.9 & 56.99 & 805101010 & 29.5 & 14.7 & 99.48\\ 
807022010 & 46.0 & 23.0 & 15.78 & 806103010 & 20.5 & 10.3 & 57.79 & 806108010 & 20.6 & 10.3 & 99.49\\ 
807020010 & 46.0 & 23.0 & 16.01 & 806139010 & 17.5 & 8.8 & 58.08 & 804061010 & 56.8 & 28.4 & 99.92\\ 
804056010 & 14.2 & 7.1 & 16.01 & 806127010 & 20.4 & 10.2 & 58.39 & 805108010 & 24.9 & 12.4 & 99.95\\ 
805103010 & 12.9 & 6.4 & 16.07 & 701007020 & 71.4 & 35.7 & 59.21 & 806144010 & 20.6 & 10.3 & 100.05\\ 
806135010 & 18.6 & 9.3 & 16.16 & 701007010 & 6.8 & 3.4 & 64.34 & 806132010 & 13.9 & 7.0 & 100.37\\ 
807019010 & 27.4 & 13.7 & 16.22 & 804066010 & 42.9 & 21.5 & 64.87 & 806121010 & 14.1 & 7.1 & 107.34\\ 
806123010 & 19.7 & 9.8 & 16.44 & 806116010 & 21.7 & 10.8 & 65.11 & 805112010 & 26.2 & 13.1 & 107.82\\ 
805046010 & 35.2 & 17.6 & 16.62 & 806104010 & 26.4 & 13.2 & 65.96 & 806109010 & 13.7 & 6.9 & 108.17\\ 
805045010 & 53.5 & 26.8 & 17.91 & 805099010 & 18.6 & 9.3 & 65.97 & 805116010 & 24.9 & 12.8 & 108.29\\ 
805047010 & 33.4 & 16.7 & 18.76 & 806140010 & 12.6 & 6.3 & 66.32 & 806145010 & 25.5 & 12.7 & 108.32\\ 
807023010 & 27.1 & 13.6 & 19.10 & 804059010 & 36.6 & 18.3 & 66.40 & 806133010 & 16.2 & 8.1 & 108.99\\ 
807021010 & 35.8 & 17.9 & 19.13 & 805106010 & 19.9 & 9.9 & 66.53 & 804069010 & 60.8 & 30.4 & 115.20\\ 
805048010 & 29.1 & 14.5 & 19.13 & 806128010 & 20.4 & 10.2 & 66.90 & 806122010 & 20.7 & 10.3 & 115.46\\ 
801049030 & 61.0 & 30.5 & 27.74 & 806117010 & 20.4 & 10.2 & 73.79 & 806110010 & 20.7 & 10.4 & 116.21\\ 
801049020 & 53.7 & 26.9 & 31.21 & 805110010 & 18.0 & 9.0 & 74.38 & 805102010 & 25.8 & 12.9 & 116.24\\ 
806112010 & 21.7 & 10.8 & 31.37 & 806105010 & 17.3 & 8.6 & 74.60 & 804062010 & 54.5 & 27.4 & 116.70\\ 
804064010 & 19.1 & 9.6 & 31.44 & 806141010 & 22.2 & 11.1 & 74.79 & 805109010 & 30.7 & 15.3 & 116.74\\ 
806100010 & 18.0 & 9.0 & 32.26 & 805114010 & 13.7 & 6.9 & 74.82 & 806146010 & 14.6 & 7.3 & 117.04\\ 
805097010 & 21.2 & 10.5 & 32.47 & 808085010 & 37.4 & 18.7 & 74.85 & 806134010 & 22.0 & 11.0 & 117.10\\ 

\hline
\hline\hline
\end{tabular}
\label{table:obs}
\end{center}
\end{table*}
}

\normalsize 
The contribution of the soft local X-ray background (including local hot bubble and galactic halo) is 
negligible in our fitting band (1.95 $-$ 6 keV), while the contribution of the cosmic X-ray background (CXB) may still be significant. To account for the contribution of CXB we add a power-law component to the model. The normalization of the power-law model is left free, while the index is fixed to 1.41 in our fits. We check for possible systematic effects regarding the CXB in Section~\ref{sec:systematics}.

\begin{table}
\begin{center}
\caption{Definitions of the used spectral extraction regions in arcmin and kpc from the cluster center. `Region 2-4' is the combination of Regions 2 through 4 (the full off-center dataset).}
\label{tab:region-def}
\begin{tabular}{lccccc}
\hline
\hline
Region & inner $d$ & outer $d$ & inner $d$ & outer $d$ \\
Name	& \it arcmin & \it arcmin & \it kpc & \it kpc \\
\hline
Region 1	&	 0 & 8.3 & 0 & 182  \\
Region 1a  & 0 & 2 & 0 & 44 \\
Region 1b  & 2 & 4.5 & 44 & 98 \\
Region 1c  & 4.5 & 8.3 & 98 & 182 \\
Region 2	& 8.3 & 25 & 182 & 545 \\
Region 3	&  25 & 40 & 545 & 873 \\
Region 4	& 40 & 130 & 873 & 2836 \\
Region 2-4 & 8.3 & 130 & 182 & 2836 \\
\hline
\hline
\end{tabular}
\end{center}
\end{table}
\begin{figure}[]
\centering
\vspace{3mm}
\includegraphics[width=8.5cm, angle=0]{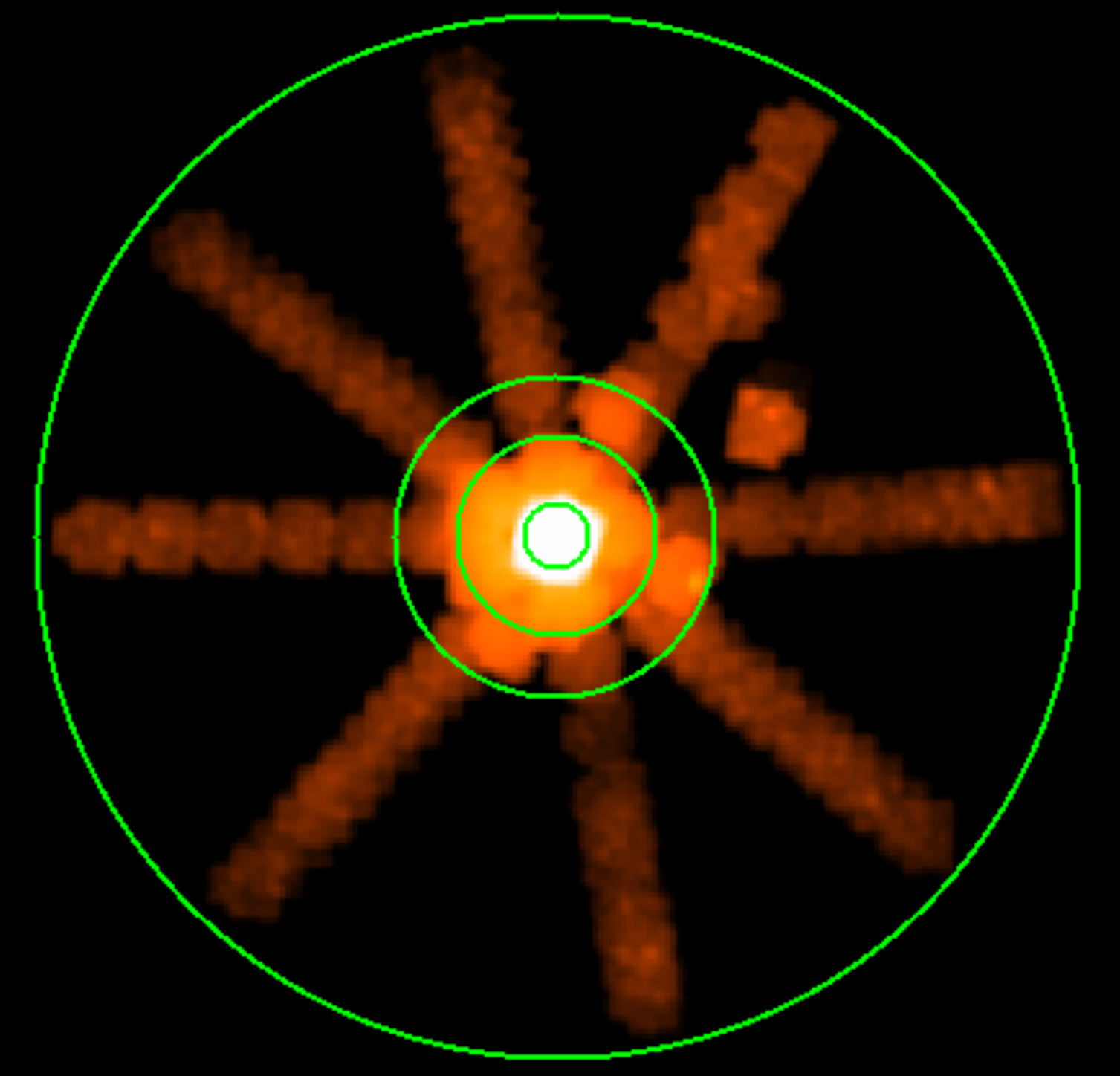}
\caption{Countmap of all pointings used in the present analysis, with radial extraction regions shown at 8.3$^\prime$, 25$^\prime$, 40$^\prime$ and 130$^\prime$.}
\label{fig:mosaic}
\end{figure}

The atomic lines and their rest-frame energies included in our model are (see also Table~\ref{tab:atomlines}): Al \textsc{xiii} (2.05 keV), Si \textsc{xiv} (2.01 keV, 2.37 keV, and 2.51 keV), Si \textsc{xiii} (2.18 keV, 2.29
keV, and 2.34 keV), S \textsc{xv}
(2.46 keV, 2.88 keV, 3.03 keV), S \textsc{xvi} (2.62 keV), Ar \textsc{xvii} (triplet at 3.12 keV, 3.62
keV, 3.68 keV), Cl {\sc XVI} (2.79 keV), Cl {\sc XVII} (2.96 keV), Cl  {\sc XVII} (3.51 keV)
 K \textsc{xviii} (triplet 3.47 keV, 3.49 kev and 3.51 keV), K \textsc{xix}
(3.71 keV), Ca \textsc{xix} (complex at 3.86 keV, 3.90 keV, 4.58 keV), Ar
\textsc{xviii} (3.31 keV, 3.93 keV), Ca \textsc{xx} (4.10 keV), Cr
\textsc{xxiii} (5.69 keV). 
After the first iteration the $\chi^2$ improvement for the inclusion of each of these lines is determined, and lines that do not improve the fit by more than a $\Delta\chi^2$ of 2 are removed from the model (on a region-by-region basis).

\begin{table}
\begin{center}
\caption{List of atomic lines and their rest-frame energies included in the model.}
\label{tab:atomlines}
\begin{tabular}{llll}
\hline
\hline
Ion & E & Ion & E \\
 & keV & & keV \\
 \hline
Al \textsc{xiii} & 2.05  & Cl  {\sc xvii} & 3.51  \\
Si \textsc{xiv} & 2.01, 2.37, 2.51  & K \textsc{xviii} & 3.47, 3.49, 3.51  \\
Si \textsc{xiii} & 2.18, 2.29, 2.34  & K \textsc{xix} & 3.71  \\ 
S \textsc{xv} & 2.46, 2.88, 3.03  & Ca \textsc{xix} & 3.86, 3.90, 4.58  \\ 
S \textsc{xvi} & 2.62, 3.28  & Ar \textsc{xviii} & 3.31, 3.93  \\
Ar \textsc{xvii} & 3.12, 3.62, 3.68  &  Ca \textsc{xx} & 4.10  \\
Cl {\sc xvi} & 2.79  & Cr\textsc{xxiii} & 5.69  \\
Cl {\sc xvii} & 2.96  & & \\
\hline
\hline
\end{tabular}
\end{center}
\end{table}

It is crucial to determine the fluxes of S \textsc{xv} at 2.46 keV and S \textsc{xvi} at 2.62 keV accurately for temperature estimation, as this line ratio is a very sensitive temperature diagnostic, especially valuable for detecting the presence of cool gas. However, the band where S \textsc{xv} and S \textsc{xvi} are located, is crowded with strong Si \textsc{xiv} lines. We therefore tie the fluxes of Si \textsc{xiv} (2.01 keV: 2.37 keV: 2.51 keV) to each other with flux ratios of (21:3.5:1). We also tie S \textsc{xv} (2.46 keV : 2.88 keV) lines with a flux ratio of (9:1). These ratios are based on the theoretical predictions for the typical temperatures we measure. The observed fluxes of some of the strong atomic lines in our fitting band are given in Table \ref{table:bestfit}.

To model the fluxes of the K {\sc xviii}, Cl {\sc xvii}, and Ar {\sc xvii} lines nearest to the 3.5 keV energy in question, we use temperature estimates indicated by other lines. The line ratios of S \textsc{xv} (1s$^1$2p$^1$ $\rightarrow$ 1s$^2$) at 2.46 keV to S \textsc{xvi} (2p$^1$ $\rightarrow$ 1s$^1$) at 2.62 keV and Ca \textsc{xix} (1s$^1$ 2p$^1$ $\rightarrow$ 1s$^2$) at 3.9 keV to Ca \textsc{xx} (2p$^1$ $\rightarrow$ 1s$^1$) at 4.11 keV are excellent temperature probes -- especially sensitive to the presence of cool gas (see \citet{Bulbul:14b} for discussion). The fluxes of lines from Cl~\textsc{xvii} and Ar~\textsc{xvii} at 3.51~keV and 3.62~keV are restricted by the other lines of the same ions detected at 2.96~keV and 3.12~keV respectively.

The emissivities of K \textsc{xviii}, K \textsc{xix}, Cl \textsc{xvii}, and Ar \textsc{xvii} lines are higher at the lower temperature ranges for  each model, which are determined from the S \textsc{xv} to S \textsc{xvi} line ratios. We use factors of 0.1 and 3 over the {\em highest values within the allowed temperature ranges} for these fluxes as lower and upper bounds for the normalizations of the Gaussian lines as described in Bu14. The factor 3 gives a conservative allowance for variation of the relative elemental abundances between the S and  K, Cl, and Ar ions.

\subsection{Systematics}
\label{sec:systematics}
In addition to the atomic model uncertainties (which we account for by using conservatively wide intervals for the allowed fluxes of the atomic lines), the main source of systematic uncertainty regarding the models is the CXB power-law component. In order to estimate the effect of this uncertainty on the other model parameters we perform the following simulations using XSPEC's \tt fakeit \rm command. 
Starting from the best-fit model, a new power-law normalization is randomly drawn uniformly from the $1\sigma$ range of the originally measured normalization.
This is repeated 1000 times, and a simulated spectrum is generated each time (with the input model only differing in power-law normalization).
The simulated spectra are refit and from the resulting population the 68\% intervals of the distribution for each free parameter are recorded. 
These are then added in quadrature to the statistical uncertainty from the best-fit model to the real data. 
The total (statistical and systematic) errors on the best-fit parameters are given in Table~\ref{table:bestfit}.
{
\begin{table}[ht!]
\begin{center}
\caption{Percentage redistribution between the inner annuli due to the effects of PSF smearing, as described in Section~\ref{sec:systematics}. Numbers represent the fraction of photons that are emitted from one annulus, and detected in another. }
\renewcommand{\arraystretch}{1.4}
\begin{tabular}{l|cccc}
\hline\hline
Region & \multicolumn{4}{c}{Region detected in} \\ 

emitted from & 0-2 & 2-4.5 & 4.5-8.3 & $>$8.3 \\ 
0-2 & 0.60 & 0.33 & 0.03 & 0.00 \\ 
2-4.5 & 0.09 & 0.68 & 0.19 & 0.01 \\ 
4.5-8.3 & 0.00 & 0.08 & 0.80 & 0.08 \\ 
$>$8.3 & 0.00 & 0.01 & 0.15 & 0.78 \\ 

\hline\hline
\end{tabular}
\label{tab:psfsmearing}
\end{center}
\end{table}
}

Due to {\it Suzaku}'s relatively large PSF, some X-ray photons that originate from one particular region on the sky may be scattered elsewhere on the detector. Since the region sizes we used in this work are similar or relatively large compared to the PSF size of the XIS mirrors, the effect is expected to be small. 
The effect of PSF spreading on the flux of the $\sim$3.5~keV line depends on its origin, and we therefore examine two scenarios. Firstly we consider the case where the flux of the line is distributed according to the broadband X-ray surface brightness as described by the higher resolution imaging of the {\it XMM Newton} PN observation of the Perseus cluster core (observation ID 0305780101). We use ray-tracing simulations of $2\times10^{6}$ photons performed through {\it xissim} \citep{Serlemitsos:2007} with our best-fit model and the PN surface brightness map as input, to determine the scattered photons per sub-region. Table~\ref{tab:psfsmearing} reports the results in terms of the fraction of photons that are emitted in one region and detected in the other. These results are consistent with the photon fractions reported in \citep{Bautz:2009} and \citep{Bulbul:15}.
The second scenario that we examine using the same methodology, is when the $\sim$3.5~keV line originates from dark matter decay and therefore follows a NFW profile. In this case, the redistribution fraction change only slightly from the ones in Table~\ref{tab:psfsmearing}, at most by a few percent-points. The dependence on the details of the NFW assumed is even smaller. 
The net effect of the PSF spreading on the measured fluxes in each regions depends more strongly on the input (or true) distribution than do the redistribution fractions. It is as follows. For the regions 1a through 1c respectively, in the case that the line follows the broadband surface brighness, the measured flux in the line would be underestimated by $\sim$31\%, overestimated by $\sim$8\% and overestimated by $\sim$22\%. In the case that the line flux follows the NFW distribution, the measurement would be underestimated by $\sim$8\%, overestimated by $\sim$3\% and overestimated by $\sim$2\%.
In Section~\ref{sec:conclusion} we will discuss the implications of this on our results, but since the origin of the line at this point is unclear, we will refrain from applying a correction for either scenario in what follows unless explicitly noted.

{
\begin{table*}[ht!]
\begin{center}
\caption{The best-fit parameters of the model. The fluxes of the S \textsc{xv}, S \textsc{xvi},Cl \textsc{xvii},  Ar \textsc{xviii} Ca \textsc{xix}, and Ca \textsc{xx} lines are in the units of $10^{-6}$ pht cm$^{-2}$ s$^{-1}$. Fields with a `-' indicate the absence of this component from the model. The $\chi^2$ reported does not include a $\sim$3.5~keV model component.}
\renewcommand{\arraystretch}{1.4}
\begin{tabular}{lcccccc}
\hline\hline
	Model  	& Reg 1 & Reg 2 & Reg 3 & Reg 4 & Reg 2-4\\
	Parameter	&  (0$^\prime$--8.3$^\prime$) & (8.3$^\prime$--25$^\prime$) & (25$^\prime$ -- 40$^\prime$) & (40$^\prime$--130$^\prime$) & (8.3$^\prime$--130$^\prime$) \\\hline
kT$_{1}$ (keV)				& 3.09 $\pm$ 0.04	& 6.52 $\pm$ 0.11 & 6.10 $\pm$ 0.29 & 5.91 $\pm$ 0.50 & 4.64 $\pm$ 0.07	 \\
N$_{1}$ (10$^{-2}$ cm$^{-5}$)	& 5.54$_{-1.33}^{+3.23}$ & 3.69 $\pm$ 0.033 & 0.57 $\pm$ 0.016 & 0.09 $\pm$ 0.005 & 0.60 $\pm$ 0.007		\\
kT$_{2}$ (keV)				& 5.78 $\pm$ 0.03	& - & - & - & - \\
N$_{2}$ (cm$^{-5}$)			& 0.54 $\pm$ 0.04 & - & - & - & - \\
Power-Law Norm (10$^{-4}$)		& 7.71 $\pm$ 0.65 & 4.62 $\pm$ 1.28 & 0.00 $\pm$ 0.40 & 0.88 $\pm$ 0.10 & 5.13 $\pm$ 0.17 \\
Flux of the S \textsc{xv} 		& 2.71 $\pm$ 0.05 $\times10^2$ & 5.60 $\pm$ 4.12 & 2.06 $\pm$ 1.85 & 0.72 $\pm$0.64 & 1.34 $\pm$ 0.85	\\
Flux of the S \textsc{xvi} 		& 7.64 $\pm$ 0.07$\times10^2$ & 23.17 $\pm$3.45 & 3.14 $\pm$ 1.62 & 1.21 $\pm$ 0.45 & 5.10 $\pm$ 0.70 \\
Flux of the Cl \textsc{xvii}		& 0.22 $\pm$ 0.04 $\times10^2$ & - & - & - & - \\
Flux of the Ar \textsc{xviii}		& 2.07 $\pm$ 0.04 $\times10^2$ & 7.35 $\pm$ 2.16 & - & 0.61 $\pm$ 0.27 & 2.11 $\pm$ 0.59 \\
Flux of the Ca \textsc{xix}		& 1.77 $_{-0.17}^{+0.34}$ $\times10^2$ & 3.96 $\pm$ 4.56 & 1.14 $\pm$ 0.98 & - & 1.07 $\pm$ 0.55	 \\
Flux of the Ca \textsc{xx}		& 1.43 $\pm$ 0.03 $\times10^2$ & 4.7 $\pm$ 1.69 & - &-  & 0.93 $\pm$ 0.39\\
$\chi^{2}$ (dof)				& 2504.4 (2170) & 2919.0 (3061) & 3276.1 (3063) & 3880.3 (3062) & 3259.0 (3060) \\

\hline\hline
\end{tabular}
\label{table:bestfit}
\end{center}
\end{table*}
}
{
\begin{table*}[ht!]
\begin{center}
\caption{Same as Table~\ref{table:bestfit}, but for the subregions of the core. The best-fit parameters of the model. The fluxes of the S \textsc{xv}, S \textsc{xvi},Cl \textsc{xvii},  Ar \textsc{xviii} Ca \textsc{xix}, and Ca \textsc{xx} lines are in the units of $10^{-4}$ pht cm$^{-2}$ s$^{-1}$. Fields with a `-' indicate the absence of this component from the model.}
\renewcommand{\arraystretch}{1.4}
\begin{tabular}{lccc}
\hline\hline
	Model  	& Reg 1a & Reg 1b & Reg 1c \\
	Parameter & (0$^\prime$--2$^\prime$) & (2$^\prime$--4.5$^\prime$) & (4.5$^\prime$--8.3$^\prime$) \\
	\hline
kT$_{1}$ (keV)				& 3.35 $\pm$ 0.11 	&  4.85 $\pm$ 0.04 & 6.41 $\pm$  0.22	 \\
N$_{1}$ (10$^{-2}$ cm$^{-5}$)	& 0.11 $\pm$ 0.03 & 0.12 $\pm$ 0.06 & 0.22 $\pm$ 0.01\\
kT$_{2}$ (keV)				& 5.72 $\pm$ 0.29	& 6.02 $\pm$ 0.24 & -   \\
N$_{2}$ (cm$^{-5}$)			& 0.16 $\pm$ 0.04 & 0.20 $\pm$ 0.03 & - \\
Power-Law Norm (10$^{-4}$)	& 4.16 $\pm$ 0.51 &  1.77$\pm$ 0.63 & 5.11 $\pm$ 0.16 \\
Flux of the S \textsc{xv} 		& 1.74 $\pm$  0.07  &  1.44   $\pm$ 0.16  &  0.65 $\pm$ 0.16  \\
Flux of the S \textsc{xvi} 		& 4.39 $\pm$ 0.07  &  4.20 $\pm$ 0.07 &   1.99 $\pm$ 0.09 \\
Flux of the Cl \textsc{xvii}		& 0.28  $\pm$ 0.06   & - & - \\
Flux of the Ar \textsc{xviii}		& 1.31  $\pm$ 0.13  & 1.19 $\pm$ 0.07 & 0.39 $\pm$  0.11 \\
Flux of the Ca \textsc{xix}		& 1.14  $\pm$ 0.12  & 1.03  $\pm$  0.04 &  0.39 $\pm$ 0.05	 \\
Flux of the Ca \textsc{xx}		& 0.71 $\pm$ 0.04  &  0.78 $\pm$ 0.05   & 0.48    $\pm$ 0.04 \\
$\chi^{2}$ (dof)				& 2317.3 (2168) & 2450.8 (2168) & 2401.7 (2168)  \\

\hline\hline
\end{tabular}
\label{table:bestfit-core}
\end{center}
\end{table*}
}%

\section{Results}
\label{sec:results}
\subsection{Perseus Center}
\label{sec:on-center}

We initially extract source and background spectra from a circular region surrounding the cluster's center with a radius of 8.3$^{\prime} $ (we refer to this region as {\it Region 1}). The total filtered on-axis FI/BI exposure times are 1.0/0.67 Ms. There are 1.4$\times10^{7}$ source counts in the background-subtracted FI spectrum and 1$\times10^{7}$ in the BI spectrum

We model the 1.95 to 6 keV band with the continuum and lines as described in the previous section (Section \ref{sec:reduction}). The best-fit values of the model are given in Table \ref{table:bestfit}. The plasma temperature measured from the continuum (3.09$\pm$0.04 keV) is in agreement with the plasma temperature estimated from the S \textsc{xv} to S \textsc{xvi} line flux ratio (3.13 keV) at a 1$\sigma$ level. We stress again that the S line ratio is very sensitive to cool gas. The peak emissivity of the  S \textsc{xv} line is at kT$\approx$ 1 keV; thus, if any significant cool gas phase were present, the line ratio temperature would be biased toward it. This plasma temperature is also in good agreement with the temperatures measured from the {\it XMM-Newton} and {\it Chandra} observations of the Perseus cluster \citep{Bulbul:14,Bulbul:14b}. 

\begin{figure}[]
\centering
\vspace{3mm}
\hspace{-4mm}\includegraphics[width=8.9cm, angle=0]{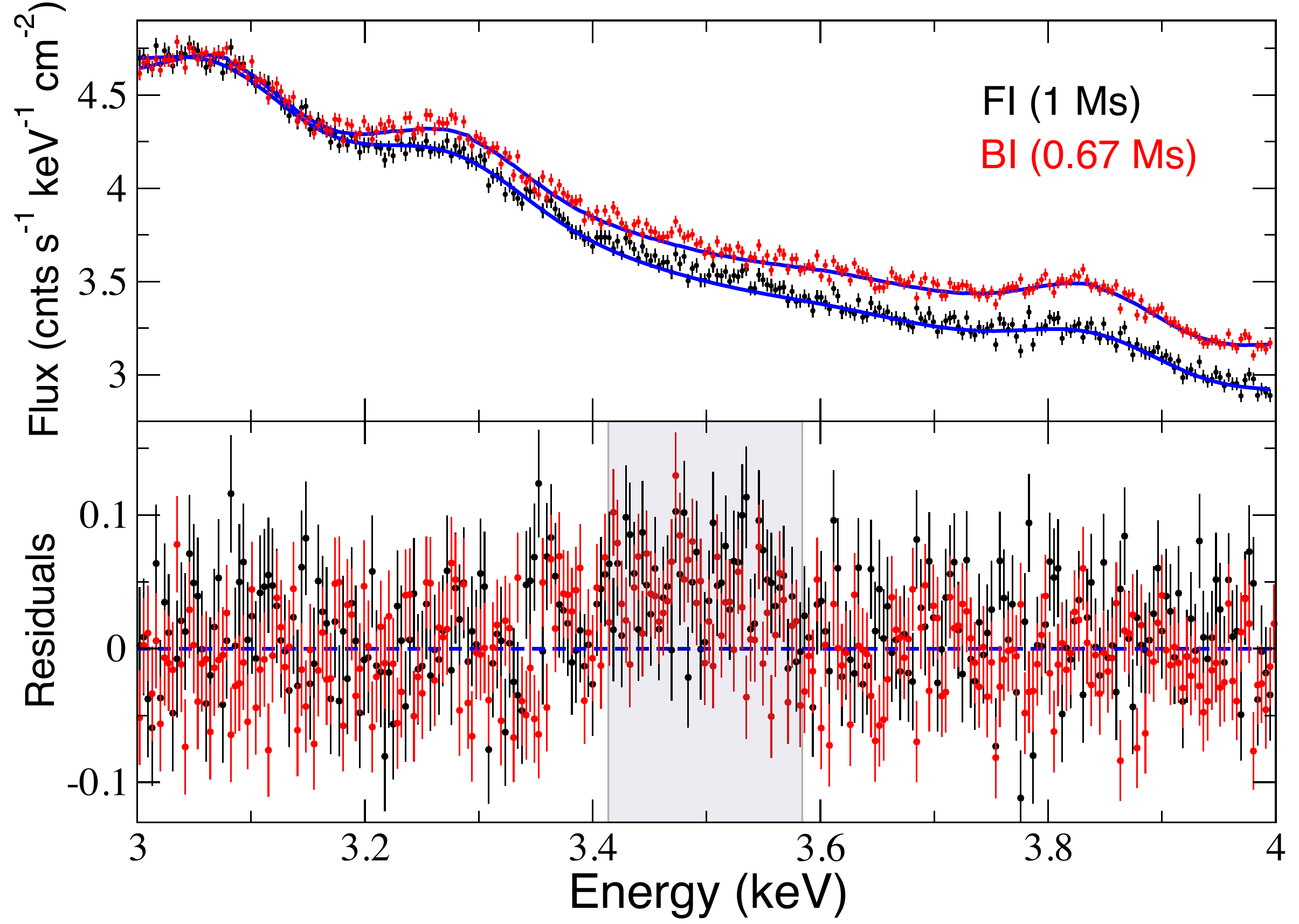}
\caption{Observed {\it Suzaku} FI and BI Spectrum of the Perseus cluster core (Region 1). The residuals around 3.5 keV (redshifted) are visible clearly (shaded area in the bottom panel). The model shown in the figure includes contributions from the nearby K {\sc xviii}, Cl {\sc xvii}, and Ar {\sc xvii} lines. The 3.5 keV rest-frame energy corresponds to 3.49 keV in this plot.}
\label{fig:corespec}
\end{figure}

Estimating the fluxes of detected lines is crucial for determining the flux around the 3.5 keV line. 
For a sanity check, we compare the intensities of the three lines from ions (i.e., Si {\sc xiv}, Ar {\sc xvii}, Cl {\sc xvii}) detected significantly in the fitting band with the estimates based on the observed S {\sc xv}~/~S {\sc xvi} line ratio. 
Si {\sc xiv} line at 2 keV is detected significantly with a flux of $(1.24\ \pm\ 0.01)\times10^{-3}$~pht~cm$^{-2}$~s$^{-1}$. 
The predicted Si {\sc xiv} flux from a $\sim3.1$ keV plasma is $1.38\times10^{-3}$~pht~cm$^{-2}$~s$^{-1}$ using AtomDB, indicating that S and Si have relative abundances of 0.9$\pm$0.01 with respect to the \citet{Asplund:2009} solar abundances.
The measured Ar {\sc xvii} at 3.12 keV is $2.07\pm0.41\times10^{-4}$~pht~cm$^{-2}$~s$^{-1}$, while the flux estimated using AtomDB is $1.30\times10^{-4}$~pht~cm$^{-2}$~s$^{-1}$. The implied abundance ratio of Ar to S is 1.6$^{+0.31}_{-0.32}$ with respect to the solar abundance. 
Unlike in the stacked {\it XMM-Newton} observations of a large sample of clusters and the {\it XMM-Newton} and {\it Chandra} observations of the Perseus cluster (from Bo14 and Bu14), we detect a very faint Cl Ly-$\alpha$ line at 2.96 keV in the {\it Suzaku} spectrum of the Perseus core. 
The measured $(2.20\ \pm\ 0.4)\times10^{-5}$~pht~cm$^{-2}$~s$^{-1}$) and estimated ($1.93\times10^{-5}$~pht~cm$^{-2}$~s$^{-1}$) Cl Ly-$\alpha$ fluxes indicate that the abundance ratio of Cl to S is $\sim$1.1$^{+0.25}_{-0.18}$ with respect to the solar abundance. The best-fit flux of the K {\sc xix} line at 3.70~keV is $6.0\pm4.0\times10^{-6}$~pht~cm$^{-2}$~s$^{-1}$. The predicted flux of the line ($3.4\times10^{-6}$~pht~cm$^{-2}$~s$^{-1}$) shows that the abundance ratio of K to S
is 1.8$\pm$1.2 with respect to solar.

{
\begin{table*}[ht!]
\begin{center}
\caption{Estimated fluxes of the Cl \textsc{xvii}, K \textsc{xviii}, Ar DR \textsc{xvii} lines are in the units of $10^{-8}$ pht cm$^{-2}$ s$^{-1}$ from AtomDB. The fluxes (and not the temperature) in this table are dependent on the assumed solar abundance \citep{Asplund:2009}, and are employed in the fits by setting the upper and lower allowed limits for the fitting procedure to 3 times and 0.1 times this flux, respectively. 
Temperature ranges implied by uncertainty of the measured lines are shown for illustrative purposes. }
\renewcommand{\arraystretch}{1.4}
\begin{tabular}{lccccccccc}
\hline\hline
Parameter  	& Reg 1 & Reg 1a & Reg 1b & Reg 1c	& Reg 2	& Reg 3	& Reg 4 & Reg 2-4	\\
				&  \\\hline
kT based on S (keV) & 3.13$\pm$0.03 & 2.97$\pm$0.06 & 3.18$\pm$0.17 & 3.25$\pm$0.36 & 3.74$^{+1.23}_{-1.69}$ & 2.37$^{+0.90}_{-2.37}$ & 2.47$^{+0.90}_{-1.56}$ & 3.60$^{+1.00}_{-1.34}$ \\ 
kT based on Ca (keV) & 4.02$\pm$0.29 & 3.65$\pm$0.16 & 3.92$\pm$0.11 & 4.85$\pm$0.36 & 4.77$^{+2.32}_{-4.77}$ & -- & -- & 4.14$^{+1.11}_{-1.36}$ \\ 
Flux of Cl~{\sc XVII} at 2.96~keV & 1932.9 & 1085.6 & 1068.9 & 510.8 & 62.2 & 6.79 & 2.70 & 13.5 \\ 
Flux of Cl~{\sc XVII} at 3.51~keV & 295.3 & 164.8 & 163.6 & 78.4 & 9.69 & 1.00 & 0.40 & 2.10 \\ 
Flux of K~{\sc XVIII} at 3.47~keV & 227.8 & 138.3 & 122.6 & 56.4 & 5.32 & 1.13 & 0.43 & 1.25 \\ 
Flux of K~{\sc XVIII} at 3.49 keV & 112.4 & 68.2 & 60.5 & 27.9 & 2.65 & 0.57 & 0.22 & 0.62 \\ 
Flux of K~{\sc XVIII} at 3.51~keV & 471.1 & 280.1 & 255.3 & 118.5 & 11.8 & 2.13 & 0.82 & 2.73 \\ 
Flux of Ar DR {\sc XVII} at 3.62~keV & 56.9 & 38.1 & 29.8 & 13.1 & 0.97 & 0.50 & 0.17 & 0.24 \\ 

\hline\hline
\end{tabular}
\label{table:estflux}
\end{center}
\end{table*}
}

To estimate the flux of the 3.5 keV line, we model the possibly contaminating K {\sc xviii} (3.47 keV: 3.49 keV: 3.51 keV), and Ar \textsc{xvii} (3.12 keV: 3.62 keV: 3.68 keV) lines with the ratios of (1: 0.5: 2.3) and (1: 1/23: 1/9).
The line ratios are estimated for the temperature indicated by the observed S {\sc xvi/xv} line ratio. We also include the Cl Ly-$\beta$ line at 3.51 keV with a flux tied to 0.15~$\times$ that of the the flux of the Cl Ly-$\alpha$ line at 2.96 keV in our fits. 
The measured best-fit K {\sc xviii} at 3.51 keV is $1.05\times 10^{-6}$~pht~cm$^{-2}$~s$^{-1}$, also in agreement with the AtomDB predictions. We note that the total flux of the K {\sc xviii} triplet between 3.47--3.51 keV is estimated at $8.11\times 10^{-6}$~pht~cm$^{-2}$~s$^{-1}$ from AtomDB (Table~\ref{table:estflux}), but that we allowed the K {\sc xviii} flux to be up to $2.5\times 10^{-5}$~pht~cm$^{-2}$~s$^{-1}$ in our fits. Additionally, we provide the flux estimates of the detected lines based on \cite{Anders:1989} solar abundance for comparison in Appendix~\ref{app:spectra-outskirts} as Table~\ref{table:estflux_AG}. In summary, the abundance ratios of detected lines implied by our measurements and AtomDB range between 1--1.7 for the strongly detected lines (including K 
{\sc xix}) in our fitting band, well within the assumed interval of a factor 0.1$-$3 regardless of assumed solar abundance sets.

Examining the 3--4 keV band in the simultaneous fits of the FI and BI observations, we find excess emission around 3.5 keV (rest energy). The residuals around 3.5 keV (which corresponds to a redshifted energy of 3.49 keV) are shown in Figure \ref{fig:corespec}. If we add a redshifted Gaussian line with energy as a free parameter, the best-fit energy of the line becomes $3.54\pm 0.01 (0.02)$ keV with a flux of $2.79_{-0.35}^{+0.35}\ (_{-0.57}^{+0.59})\times10^{-5}$ pht cm$^{-2}$ s$^{-1}$. The fit improves by $\Delta\chi^{2}$ of 62.6 for 2 degrees-of-freedom (d.o.f.), corresponding to a $\sim7.6\sigma$ detection. 

To investigate the radial behavior of the signal in the core, we divided the core into three spectral extraction regions: circular regions with radii of 0$-$2$^{\prime}$, $2^{\prime}-4.5^{\prime}$, and $4.5^{\prime}-8.3^{\prime}$. The best-fit model parameters of the {\it line-free apec} model is given in Table \ref{table:estflux}. Following the same fitting procedure described above, we find that the best-fit energy and flux of the line in the innermost $0-2^{\prime}$ region are $3.51\pm\ 0.02\ (0.03)$ keV and $9.28^{+2.62}_{-2.67} \ (^{+4.41}_{-4.33})\times10^{-6}$ pht cm$^{-2}$ s$^{-1}$. The change in the $\Delta\chi^{2}$ is 12.1 for the extra 2 d.o.f. 
In the intermediate $2^{\prime}-4.5^{\prime}$ region, the line energy is detected at $3.55\pm 0.02 \ (0.03)$ keV with a flux  of $1.67^{+0.29}_{-0.30} \ (^{+0.52}_{-0.48})\times10^{-5}$ pht cm$^{-2}$ s$^{-1}$ ($\Delta\chi^{2}$=23.3 with additional two d.o.f.).
The line is also detected  in the last $4.5^{\prime}-8.3^{\prime}$ region  at an energy of $3.58\pm 0.02 \ (0.03)$ keV with a flux of $1.61^{+0.32}_{-0.34}\ (^{+0.51}_{-0.49})\times10^{-5}$ pht cm$^{-2}$ s$^{-1}$ ($\Delta\chi^{2}$=16.5 for additional 2 d.o.f.). 
The radial profile of this signal has also been studied by \cite{Urban:14} in two spectral regions. Our results are in broad agreement once the sizes and shapes of the spectral extraction regions are taken into account, as we will discuss in Sections~\ref{sec:discussion}.

We then fit these spectra with a Gaussian model with the line energy fixed at 3.54 keV, which is the best-fit value detected in the 0--8.3$^\prime$ region. We find that the flux of the line becomes $6.54\pm 2.62\ (4.3)\ \times 10^{-6}$ pht cm$^{-2}$ s$^{-1}$ in the innermost 0--2$^\prime$ region, with a change in the $\Delta\chi^{2}$=6.23 for an additional 1 d.o.f. The flux remains the same ($1.67_{-0.28}^{+0.31}\ (_{-0.47}^{+0.49})\ \times10^{-5}$ pht cm$^{-2}$ s$^{-1}$) within the intermediate 2$^\prime$ -- 4.5$^\prime$, while the change in the $\chi^{2}$ becomes 25.9 for an additional 1 d.o.f. In the last region the line is detected with a flux of $1.27_{-0.34}^{+0.29}\ (_{-0.47}^{+0.41})\ \times10^{-5}$ pht cm$^{-2}$ s$^{-1}$ with a $\Delta\chi^{2}$ of 10.8 for additional 1 d.o.f.
The $\sim$3.5 keV line is detected with a confidence of $>3\sigma$ in all three regions within the core of the Perseus cluster.
Table~\ref{tab:detections} summarizes the above results.

\begin{table*}[ht!]
\begin{center}
\caption{Best-fit values for detected excess emission around 3.5 keV (rest frame) for the core regions. Also included is the best-fit flux in the case that the energy is fixed to the best fit from Region 1 (ie, 1 additional degree-of-freedom instead of 2). Total $\chi^2$ values are shown before the $\sim$3.5~keV line is added to the model.}
\label{tab:detections}
\renewcommand{\arraystretch}{1.6}
\begin{tabular}{llcccc}
\hline
\hline
Region    &  & Restframe E & Flux & $\Delta\chi^2$ & $\chi^2$ (dof) \\
 & & keV & $10^{-5}$ ph s$^{-1}$ cm$^{-2}$ & & \\
\hline
Region 1  & (0$^\prime$--8.3$^\prime$) & $3.54\pm 0.01 (0.02)$ & $2.79_{-0.35}^{+0.35}\ (_{-0.57}^{+0.59})$ & 62.6 & 2441.7 (2168)\\
Region 1a & (0$^\prime$--2$^\prime$) & $3.51\pm\ 0.02\ (0.03)$ & $0.93^{+0.26}_{-0.27} \ (^{+0.44}_{-0.43})$ & 12.1 & 2317.3 (2168)  \\
 & & 3.54 & $0.65\pm 0.26\ (0.43)$ & 6.23 & \\
Region 1b & (2$^\prime$--4.5$^\prime$) & $3.5\pm 0.02 \ (0.03)$ & $1.67^{+0.29}_{-0.30} \ (^{+0.52}_{-0.48})$ & 23.3 & 2450.8 (2168) \\
 & & 3.54 & $1.67_{-0.28}^{+0.31}\ (_{-0.47}^{+0.49})$ & 25.9 & \\
Region 1c & (4.5$^\prime$--8.3$^\prime$) & $3.58\pm 0.02 \ (0.03)$ & $1.61^{+0.32}_{-0.34}\ (^{+0.51}_{-0.49})$ & 16.5 & 2401.7 (2168)\\
 & & 3.54 & $1.27_{-0.34}^{+0.29}\ (_{-0.47}^{+0.41})$ & 10.8 & \\
\hline
\hline
\end{tabular}
\end{center}
\end{table*}
\subsection{Perseus Outskirts}
\label{sec:off-center}
A total of 100 \suz observations of the Perseus cluster with the nominal pointing further than 14$^\prime$ from the cluster center were retrieved from the archives, for a total cleaned FI/BI exposure of 2.72/1.36 Ms and background-subtracted source counts of 6.3$\times$10$^5$ and 4.3$\times$10$^5$. 
We divide this data into three annular spectral extraction regions. 
The first annulus (called `Region 2') starts at 8.3$^\prime$, where the central analysis of Section~\ref{sec:on-center} ends, and extends to 25$^\prime$.
`Region 3' is an annular extraction region with inner radius 25$^\prime$, and outer radius 40$^\prime$.
While the outermost annulus does not have an outer radius imposed, the outermost pointing is centered on 117$^{\prime}$  from the Perseus cluster core, so that all data used in this study comes from within 130$^\prime$. This is `Region 4' in Table~\ref{tab:region-def}. The same table contains the sizes of all regions in angular and physical scales. A visual representation is given in Figure~\ref{fig:mosaic}. 
As will become apparent in later sections, it is also useful to create a single stacked dataset of all these off-center observations in order to obtain better statistics. This is referred to as `Region 2-4' in Table~\ref{tab:region-def}.

To further obtain maximum photon statistics, in the results reported here for the off-center data, no point sources were removed. A parallel analysis of a version of the dataset with the point sources removed as detected by \cite{Urban:14}, did not reveal large qualitative differences. Since we have not detected the 3.5~keV line in the outskirts, we only show the higher-statistics dataset that did not mask the point sources.

The spectral modeling of the off-center is performed as described in Section~\ref{sec:reduction}, unless noted otherwise. 
The energy band used for fitting the off-center observations is reduced to 1.95 -- 5.7 to avoid a strong negative residual in the XIS 1 spectra. This is likely associated with an imperfect background subtraction of the instrumental Mn-K$\alpha$ line (see also \cite{Sekiya:2015jsa}).
In addition to the tied line ratios mentioned in Section~\ref{sec:reduction}, the off-center analysis also tied the flux of the S~{\sc XV} line at 3.03 to S~{\sc XV} at 2.46 with the theoretical ratio (1:40). 

As in the analysis of the central region, we utilize the observed line ratios of S and Ca where available to determine the maximum contribution of the Ar and K lines near 3.5~keV. 
The measured line ratios in most regions imply a second thermal component at somewhat lower temperature, but none of the broadband fits prefer a model with two plasma continuum components. 
As we noted in the previous section, this is not entirely unexpected for a multi-temperature environment as the broad-band fit is mostly sensitive to high temperatures and the power-law normalization of the CXB component, while the emissivity of the S lines peaks at low temperatures and thereby causes the S line ratios to be sensitive to the low temperature components. 
Therefore we modify the previously obtained models by setting the maximum allowed range for the line normalizations for the Ar, K and Cl lines around 3.5~keV to 3$\times$ the maximum shown in Table~\ref{table:estflux} indicated by the S and Ca ratios, and refitting. 

We obtained acceptable fits to the data of all off-center regions with a reduced-$\chi^2$ of around 1, except for Region 4 (the outer region), where $\bar{\chi}^2\sim 1.25$. 
This is most likely due to large radial extent of this region of the cluster that is stacked, making the single model fit insufficient.
The results of the fits of the off-center regions are shown in table~\ref{table:bestfit}. Plasma temperatures and normalizations are generally consistent with the measurements performed by \citet{Urban:13}.
However, the relatively low best-fit temperature for Region 2-4 is mainly caused by a preference for a relatively high normalization of the powerlaw. Fixing the powerlaw normalization to a lower value more in line with the outer regions, brings the temperature of the continuum component up again to above 6~keV. However, the fit with the fixed powerlaw normalization provides a worse fit by a $\Delta\chi^2$ of about 15. 
The fit otherwise shows no qualitative differences, and therefore we continue to employ the better fitting model (with fitted powerlaw normalization). As mentioned above, the best-fit continuum temperature is not used for the estimates of line strengths, rather the line ratios of well-measured S- and Ca- lines are.

With these final models in hand, we look for the presence of excess emission by adding a redshifted Guassian line component to the model at different restframe energies around $\sim$3.5~keV while leaving the normalization free. The plasma temperature and the normalizations of all other model components are left free in these fits. 
There is not a single region of the Perseus cluster outskirts for which we see significant positive line-like residuals anywhere in the vicinity of 3.5~keV (restframe). Note that none of the Ar, Cl or K lines near 3.5~keV are detected in these datasets either (i.e., contributions from these lines were allowed in the earlier fitting process described in Section~\ref{sec:reduction}, but were not required by the fits).

Not having found significant line-like residuals around 3.5~keV, we compute the flux limit for such a line for each off-center spectrum in the following way. Starting with the best-fit model we add one redshifted Gaussian at rest-frame 3.54 keV (the nominal detected value in Region 1), and vary its normalization until the new $\Delta\chi^2$ is higher by 4.0, which corresponds to a $2\sigma$ limit for a single added degree of freedom. 
The normalizations of all model components are left free, as is the plasma temperature. 
The obtained flux limits will be discussed in Section~\ref{sec:fluxprofile}.

\section{Discussion}
\label{sec:discussion}

\subsection{Line Flux and Dark Matter Profiles}
\label{sec:fluxprofile}

\begin{figure}
\hspace{-0.4cm}\includegraphics[width=8.5cm]{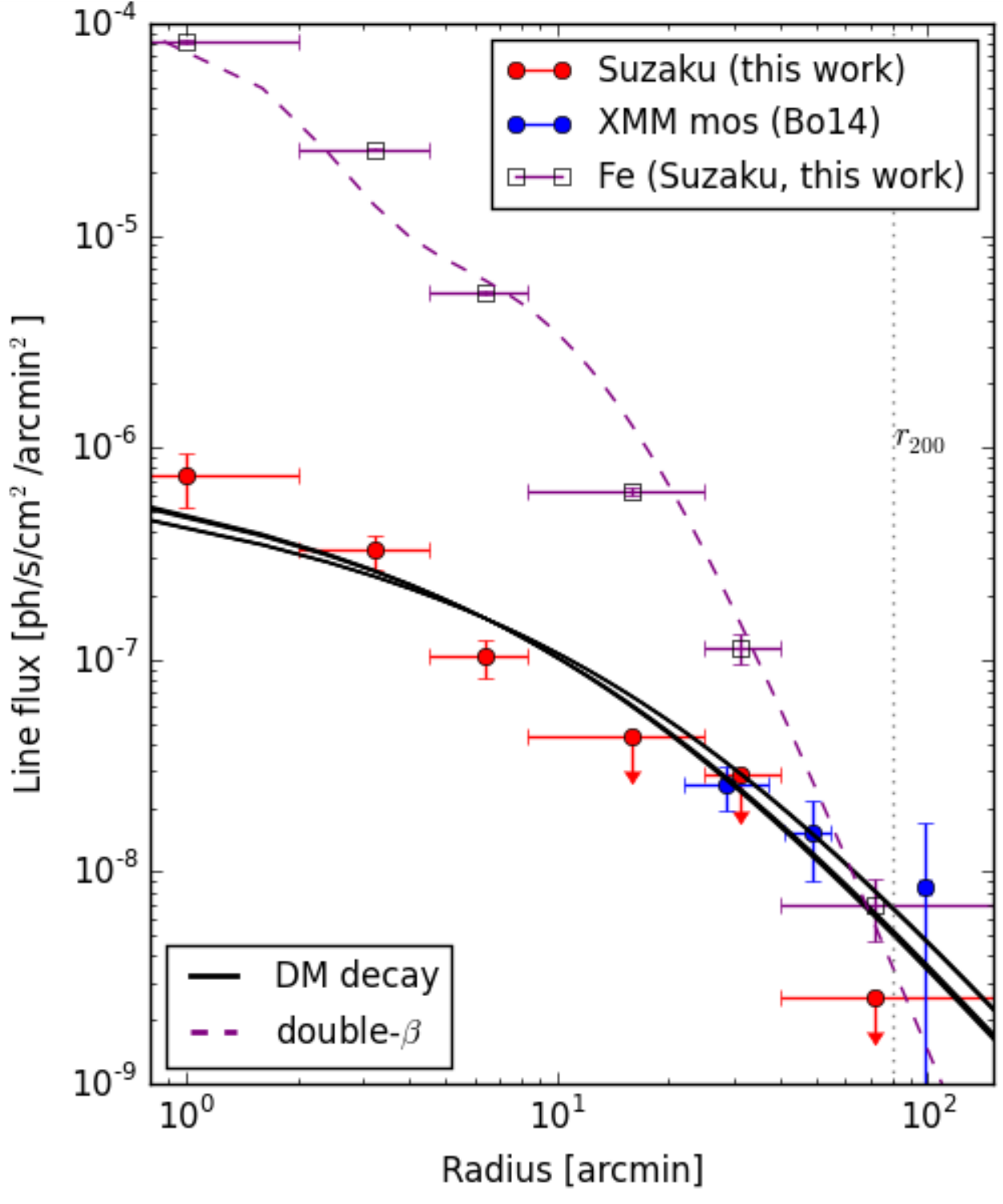}
\caption{Radial profile of the measured $\sim$3.5~keV surface brightness ($1\sigma$ error bars) and upper $2\sigma$ limits obtained from our Suzaku measurements (red), compared to the measurements of Bo14 using \xmm (blue). Black curves indicate the expected surface brightness profiles of a dark matter decay signal based on several NFW literature profiles for the dark matter distribution (see text). The normalization of these predictions is degenerate with the particle lifetime, and the shown curves have an arbitrary normalization assigned for visual purposes in this figure.
Horizontal error bars show the bracket of radial extraction regions per bin, while the central value is the dark matter column density-weighted average radius for that radial bin.
For comparison, the purple empty squares indicate measurements of the Fe~{\sc xxv} K-$\alpha$ emission at 6.7~keV in our data and the purple dashed curve shows a surface brightness profile based on the double-$\beta$ profile measured by \cite{Churazov:2003hr} but with arbitrary normalization.
Note that none of the lines shown in this figure are fitted.}
\label{fig:radial-flux}
\end{figure}

We compare our results to the behaviour expected from dark matter decay in this Section. For a first look, Figure~\ref{fig:radial-flux} shows the radial dependence of the surface brightness of the $\sim$3.5~keV signal. The results from this work and those obtained by Bo14 are shown in red and blue respectively. Downward pointing arrows indicate the 2$\sigma$ upper limits from the analysis of the outskirts. Expected dark matter decay signal strength for different NFW dark matter distributions (see below) is depicted by the set of black curves. It is important to note that the normalization of the expected decay signal depends on the dark matter particle lifetime and is therefore completely degenerate with the absolute mass scale of the NFW profiles. The figure shows arbitrary individual normalizations chosen to facilitate visual comparison in this case. 

Additionally, Figure~\ref{fig:radial-flux} shows the detected surface brightness of the  Fe~{\sc xxv} K-$\alpha$ line at 6.7~keV from all our \suz regions with the open purple squares as an indicative visual example of possible emission line-like behaviour. This behaviour is typically described by a double-$\beta$ profile, which is shown as the purple dashed line with parameters from \cite{Churazov:2003hr} albeit with arbitrary overall normalization in order to roughly line up with the Fe measurements. The measurements of the Fe lines and the double-$\beta$ profile are compatible with each other while showing quite a contrast with both the $\sim$3.5~keV measurements and the DM decay-like profiles. 

It is important to note that the radial behaviour as shown in this figure does not accurately reflect the effects of the varying pointings nor of the varying field-of-view shapes and sizes that are averaged in each datapoint, which will be handled in detail in the following. 

Are our non-detections in the Perseus outskirts inconsistent with the dark matter decay origin of the 3.5~keV line? In order to determine this, we compare the measurements to the predictions in the most direct way, by computing the effective dark matter mass in the field of view for each dataset. For a given field of view, this quantity depends only on the dark matter profile assumed, and is directly related to the expected signal by the particle lifetime. It is computed as follows.
For the off-center \suz data, where the different observations have been separated into concentric annuli, we divide the available pixels for a particular observation and extraction region into 25 spatial bins. 
Then we compute the dark matter column density at the center of each of those bins, given an NFW model, before converting to mass inside the effective field-of-view using the effective sky area. The exposure weighted average mass is then obtained for each region.
For the on-axis observations, the extraction regions are of a more convenient shape, allowing us to simply compute the enclosed mass within a certain projected radius for a given NFW profile.

\begin{figure*}
\includegraphics[width=\textwidth]{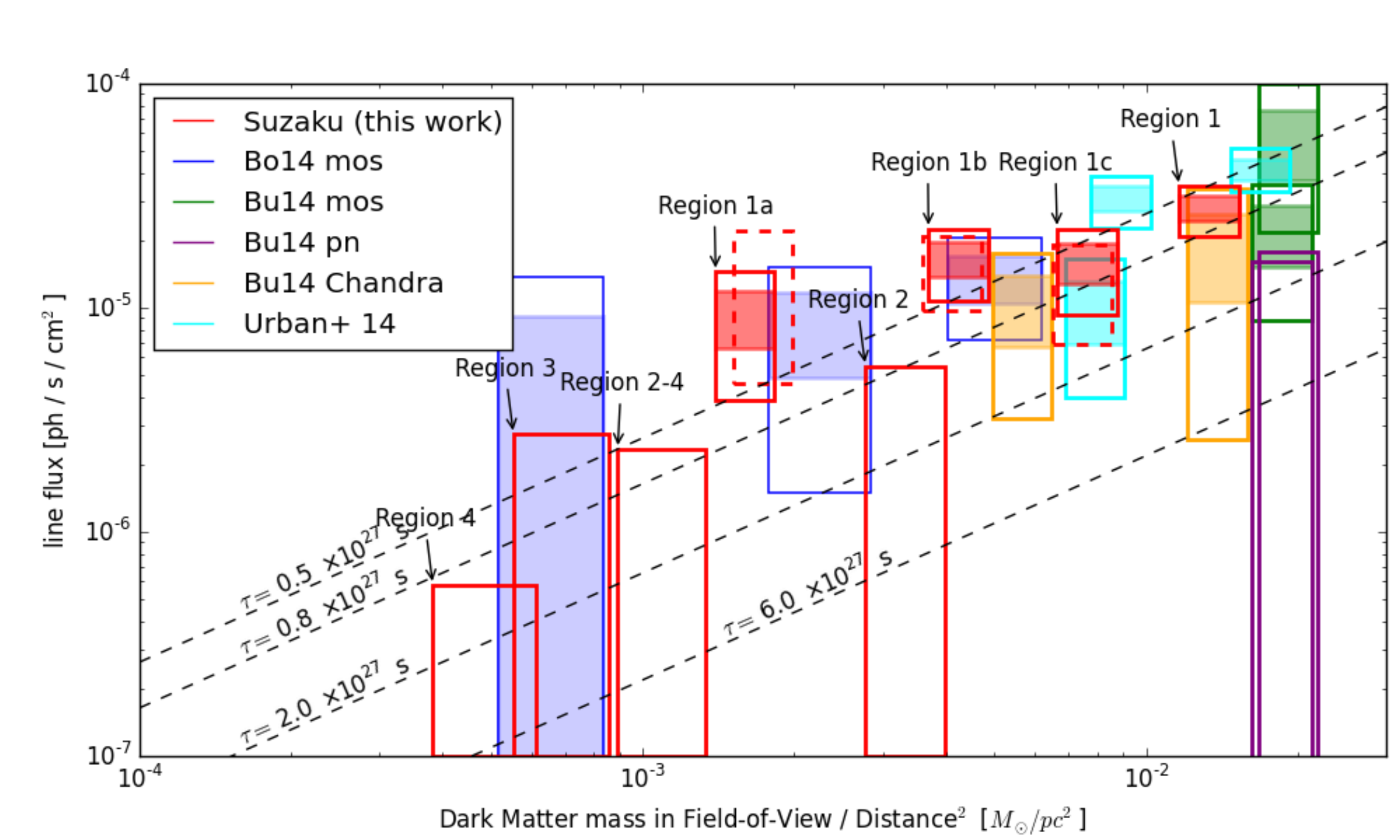}
\caption{The 3.5~keV flux as a function of the dark matter mass in the field of view. Measured by \suz in the red boxes (this work), by \xmm MOS from Bo14 in blue and from Bu14 in green, by \xmm PN in purple (Bu14), by \chan in orange (Bu14). Also shown in cyan are the \suz measurements of \cite{Urban:14} with from left to right their 'confining', 'core' and full extraction regions (see text).
Filled boxes indicate $1\sigma$ flux measurements, open boxes the $2\sigma$ interval. Boxes without a filled part and touching the x-axis indicate upper limits ($2\sigma$ for this work, reported 90\% for Bu14 pn), ie., Bu14 pn, and Regions `2', `3', `4' and `2-4'.
The dashed red boxes indicate 2$\sigma$ intervals of the \suz core subregions that have been corrected for PSF scattering using an alternative scenario for its estimation (see Sections~\ref{sec:systematics} and~\ref{sec:conclusion}).
The width of the boxes is given by the bracket of different literature NFW profiles (see text). 
Lines of constant dark matter particle lifetime are the black lines with decay rates given in the annotation. \\
{\bf NB}: this study does not constrain the value of $\tau$ as this requires and absolute mass scale to be established; the values shown are for indicative purposes. The study by \cite{Boyarsky:14b} compares different objects to this end, and uses a broader mass bracket for the Perseus cluster due to the inclusion of additional different probes of the cluster mass, extending the brackets out to longer lifetimes of order $\tau \sim 6\times10^{27}$ (see Section~\ref{sec:fluxprofile} for discussion).}
 \label{fig:xmm-suz-Mproj-flux}
\end{figure*}

We compare the results of this work with the results obtained in Bo14, Bu14 and \cite{Urban:14}. The effective dark matter mass for these observations is obtained in a similar fashion as described above.
Figure~\ref{fig:xmm-suz-Mproj-flux} shows the flux (detections and upper limits) of the $\sim$3.5~keV line as a function of dark matter mass in the field of view for a bracket of literature mass profiles. 
The red boxes marked \suz are the detections and the upper limits from this work (upper limits defined as $\Delta\chi^2$ of 4.0, or $2\sigma$ for 1 degrees of freedom). 
Lines of constant dark matter particle lifetime are shown as diagonal black lines. 
Each box represents a different spectral extraction region, for which the DM mass in that particular field of view has been computed by the method described above. This is done for three literature profiles for the Perseus cluster \citep[e.g.,][]{Simionescu:2012a,Sanchez-conde:2011,Storm:2012ty}. \cite{Storm:2012ty} makes use of the measurement of $M_{500}$ of \cite{Chen:2007sz}, determines NFW parameters through scaling relations and finally corrects for the gas fraction to get to the dark matter distribution. \cite{Sanchez-conde:2011} employs the measurement of $M_{200}$ from \cite{Reiprich:2001zv} and the scaling relation from \cite{Duffy:2008pz}. Lastly, \cite{Simionescu:2012a} derives an NFW profile for the total mass distribution directly by fitting to piecewise annular X-ray data.
The latter two do not quote dark matter only profiles, so we take the baryon fraction into account using the functional form $f_{gas} \sim r^{0.43}$ \citep{Mantz:2014} calibrated to the reported gas fraction of Perseus by \cite{Simionescu:2012a}.
Included in the bracket of computed enclosed dark matter mass are the statistical $1\sigma$ uncertainties reported in those works, although the scatter between the different profiles is larger than the statistical errors on each. 
In all computations of the enclosed dark matter mass, the different background cosmologies and differences in the definition of the NFW used in those studies have been take into account.

Here we take the effects of PSF smearing described in Section~\ref{sec:systematics} into account in the following way. As was noted, this effect is only relevant for the smaller regions 1a through 1c, and it is dependent on the origin of the $\sim$3.5~keV line. In Figure~\ref{fig:xmm-suz-Mproj-flux}, we compare the measured flux to the expected flux for a dark matter decay scenario, and we therefore apply the estimated effects of this scenario to the boxes for Regions 1a through 1c (we repeat for convenience; -8\%, +3\%, +2\% respectively). For completeness we also show the corrections for the scenario when the $\sim$3.5~keV line follows the broadband X-ray surface brightness (as described in Section~\ref{sec:systematics}, -31\%, +8\%, +22\%) as the dashed open red boxes.
In both cases, this systematic effect was also added in quadrature to the error estimate to account for the uncertainty on this effect itself.
Qualitatively, our conclusion is independent of the approach to PSF smearing used.

Not all data in Figure~\ref{fig:xmm-suz-Mproj-flux} is statistically independent. Regarding the current work (red boxes), `Region 2-4' is a compound of `Region 2', `Region 3' and `Region 4'. Regions 1a--c are subdivisions of `Region 1'. Bu14 reported 2 measurements for each of the mos (green boxes) and pn detectors (purple boxes), the difference being the excision of the central 1$^\prime$ of the Perseus cluster (the data with the core excluded is the datapoint with the lower effective dark matter mass). Their \chan measurements (yellow boxes) refer to the ACIS-S and ACIS-I chips of which the latter has the larger field-of-view and therefore higher effective dark matter mass. The 3 measurements shown of \cite{Urban:14} (cyan boxes), from right to left (higher to lower effective dark matter mass), refers to their full extraction region (full \suz field-of-view on-center), the core of the Perseus cluster (inner 6$^\prime$) and the `confining' region (full field-of-view excluding the 6$^\prime$ core). In addition, the \cite{Urban:14} study is based on the same archival data as our `Region 1' (and its sub-divided annuli). The Bo14 and Bu14 mos data from \xmm are in fact from different independent pointings. 

Our results as shown in Figure~\ref{fig:xmm-suz-Mproj-flux} indicate that the measurements and upper limits obtained with \suz in this work are internally mostly consistent with a decaying dark matter interpretation and with previous measurements. 
However, the non-detection in the outer-most region (`Region 4') is somewhat at odds with the fluxes of the measurements of the inner 2$^\prime$ (`Region 1a') and the annulus between 2$^\prime$ and 4.5$^\prime$ (`Region 1b'). Here we note that `Region 4' has the worst fit quality of all off-center datasets at a reduced-$\chi^2$ of $\sim$1.25 and the upper limit may be affected by this. 
In addition, the limit from Region 2 is marginally inconsistent with the detection in `Region 1b'. 

The very core of the Perseus cluster exhibiting relatively high $\sim$3.5~keV flux as reported in previous works is confirmed in our \suz data, but the inconsistency is less than $3\sigma$ even in the most extreme case. In addition, this enhanced flux is confined to a region smaller than $\sim$100~kpc (or $\sim$4.5$^\prime$), a large fraction of which is occupied by the brightest cluster galaxy NGC 1275, and which is well inside the cool-core. This may influence both the spectral modeling and the dark matter distribution. 
Lastly, relaxing our conservative bounds (defined as $\Delta\chi^2$ of $4.0$ for a fixed line energy) on the non-detections will alleviate the above inconsistencies.

The NFW profiles implemented in our calculations are taken from the literature as reported, all of which are based on X-ray measurements. \cite{Boyarsky:14b} uses additional literature profiles obtained by different methods for the comparison between different objects, whereas this work is concerned with the internal behaviour of the signal within the Perseus cluster only. 
Extending the mass bracket to include all of the profiles used in \cite{Boyarsky:14b} (not shown), $\tau =$ 6 $\times10^{27}$ s becomes consistent with almost all measurements. 
We stress again that absolute mass calibration is degenerate with dark matter particle lifetime $\tau$ and that this work therefore does not constrain the latter.

\subsection{Discussion of Perseus' Morphological and Dynamical State}
The use of an NFW profile for the dark matter distribution of the Perseus cluster is justified, as the cluster is reported to be a relatively relaxed cluster with a regular morphology and a moderately strong cool core \citep{Simionescu:2012a,Simionescu:2012b}. 
These studies find that even if the assumptions of spherical symmetry and hydrostatic equilibrium are relaxed to account for some evidence of gas clumping \citep{Simionescu:2012a}, their results remain consistent. 
In addition, the in- or exclusion of data from additional instruments, nor a change to a generalized NFW profile influence those results.
\cite{Simionescu:2012b} do report evidence of a past minor merger, indicated by a spiral-pattern of enhanced surface brightness across the extent of the Perseus cluster in \suz data due to gas sloshing. The infall trajectory has been determined as east-west, although the inclination is ill-constrained other than excluding edge-on. 
The initial NFW profile determined by \cite{Simionescu:2012a} was based on observations of the North-West-arm of the \suz survey of the Perseus cluster. This arm does not exhibit any evidence of this minor merger, so it is safe to conclude that for the current work it is not required to allow for any additional uncertainty in the mass profile of the Perseus cluster to account for dynamical disturbance, or irregular morphology.

\subsection{Literature Comparison}
\label{sec:discussion-urban}
The data of the Perseus core from the \suz archives employed in this work was also used by \cite{Urban:14} and \cite{Tamura:2014mta}. These works contain contradictory results, with \cite{Tamura:2014mta} not reporting any excess flux around $\sim$3.5~keV. Our results agree with the work of \cite{Urban:14} regarding the Perseus cluster. Although our extraction regions and the spectral modeling are different, the $\sim$3.5~keV line surface brightness is consistent once the different spectral extraction regions are taken into account (as can be seen in Figure~\ref{fig:xmm-suz-Mproj-flux}).

The work by \cite{Tamura:2014mta} is unable to detect the putative feature at $\sim$3.5~keV in the same data as employed in the present work and by \cite{Urban:14} even though we employ the same calibration modifications (see Section~\ref{sec:reduction}) as \cite{Tamura:2014mta}. The authors claim that the $\sim$3.5 keV line detection could be an artifact of the degeneracy
between the atomic lines and the continuum during fitting. They illustrate their claim
with an example (in their section 4.2 and figure 14), where they fit the data between 3 -- 4.2 keV with a model consisting of the plasma continuum and nine additional emission lines. Removing one of the lines from this model reveals a positive line-like residual, by design. There are a number of issues with this particular approach. Firstly, their fitting band
is too narrow to determine the continuum level accurately, and in addition, they cover their 
entire energy range with extra gaussian lines, practically guaranteeing complete degeneracy 
between line fluxes and continuum level given the large resolution of XIS detectors. 
Secondly, the lines that are added are given fluxes that are unphysically high, namely 0.2 times the 3.1 keV Ar line, whereas our Table 6 shows that that these lines are expected to be about 10 times lower (0.03 -- 0.04 
times the 3.1 keV Ar flux) than that. These fluxes were not allowed to vary and forced to be overestimated in their fit. This forces their continuum level to be underestimated, again guaranteeing that the removal of one gaussian 
model component reveals a line-like residual. A possible way to test this would be to compare the plasma temperature estimates, however, the plasma temperatures are not provided in the relevant section. 

Additionally, the line modeling in \cite{Tamura:2014mta} is less exhaustive than in our work (their 9 atomic 
lines compared to our 29). The limited number of lines used in their analysis leads to a large reduced chi-square value of 1.72 (compared to our 1.1). Indeed, most of the line emission does not get modeled properly and leads to residuals that are larger than the putative feature we detected in the fitting band. We reiterate that the putative feature is only a 1\% flux feature over the continuum and that the continuum should be modeled at that level or better to be able to detect the line. We agree that the quality of the spectral modeling is essential to our work, and that at CCD resolution one has to be very careful of the interplay between atomic lines and the
continuum. Our modeling procedure is as thorough as it is,
taking the widest possible energy range to help determine the continuum level, providing
physically motivated modeling of the atomic lines, and cross-checking the best fit line fluxes
with atomic data.

\section{Conclusion}
\label{sec:conclusion}
We have studied all available data from the \suz telescope of the Perseus cluster out to almost 1.5$r_{200}$ with the aim to investigate the radial behavior of the still unidentified line feature around 3.5~keV that was first reported in Bu14 and Bo14. We have studied the possibility that the detected 3.5~keV feature in the center of the Perseus cluster is due to atomic emission from highly-ionized nearby Ar {\sc xvii}, Cl Ly-$\beta$, and K {\sc xviii} lines in the spectral neighborhood. We detect, for the first time, Cl Ly-$\alpha$ line at 2.96 keV in clusters of galaxies, whose flux is used to calculate the flux contribution of  Cl Ly-$\beta$ line at 3.5 keV. Using measurements of various detected strong emission lines in other energy bands of the spectrum to estimate the plasma temperature and allowing for a conservatively large range of elemental abundances, we find that the 3.5~keV flux is in excess of what is allowed for atomic line emission.
We report a detection of this line feature from the central observations of the Perseus cluster with a measured flux in agreement with the previously reported detection \citep{Urban:14}. 

The \suz observations of the cluster's outskirts do not exhibit an excess of flux around 3.5~keV, nor in radially separated annular regions. The upper limits provided by the co-added outskirts observations are consistent with the dark matter decay interpretation for the origin of the signal from the Perseus cluster. 
Of course, our results are also consistent with some unknown astrophysical line originating predominantly in the dense gas of the Perseus core.

Considering the current body of work, it is not presently possible to prove conclusively the origin of the 3.5~keV line as sourced by any one process. 
The measurements in this study indicate that cluster outskirts or other low-density environments are promising targets in terms of constraining power for future observational work provided the exposure reaches deep enough. 
The most likely immediate-future gain is through employing next-generation micro-calorimeters on board the planned {\it Micro-X} \citep{Figueroa-Feliciano:2015gwa} mission, or on board {\it Hitomi} \citep{Kitayama:2014fda} if the satellite or any data thereof can be salvaged. These instruments have the energy resolution required to improve the spectral modeling, in particular with regards to the measurements of the various line emission. 
Alternative methodologies relying on different observables to distinguish dark matter decay from astrophysical or instrumental effects also offer promising possibilities. \cite{Zandanel:2015} for example suggests that the upcoming eROSITA survey \citep{Merloni:2012} will be able to distinguish dark matter decay by its behavior in an all-sky angular correlation analysis. 
Micro-calorimeters may also be able to detect the velocity shift and velocity broadening of X-ray spectral lines, which behave differently for dark matter decay or plasma emission due to the difference in dynamics between dark matter and gas, as described by \cite{Speckhard:2016}.

\section*{Acknowledgements}
The authors thank Ondrej Urban for kindly sharing with us the coordinates of point sources detected in the {\it Suzaku} field-of-view;  Larry David, Stefano Ettori, and Felipe Andrade-Santos for providing useful suggestions. 
The work of J. F. was supported by the De Sitter program at Leiden University with funds from NWO. This research is part of the Fundamentals of Science program at Leiden University.
E. B. acknowledges support by NASA through contracts NNX14AF78G and NNX123AE77G. The work of R. S. was funded in part by NASA Grant NNX15AE16G. The work of D. I. was supported by a research grant from VILLUM FONDEN.

\appendix
\section{Comment on ``Discovery of a 3.5 keV line in the Galactic Centre and a critical look at the origin of the line across astronomical targets"}

\cite{Jeltema:14} presented an analysis stating that if using a multi-component plasma and summing the fluxes from those using calcium to estimate the flux, the high temperature component would dominate, leading to a potentially large underestimate of the K~{\sc xviii} triplet flux.

However, in our analysis we calculated the emissivity of the K {\sc xviii} triplet based on fluxes from both Ca and S, one of which peaks at a higher temperature, and one which peaks at a lower temperature. We have allowed for three times the maximum K~{\sc xviii} flux permitted by either of these emissivity estimates as a safety margin. Crucially, the use of the S  {\sc xvi} emissivity, and not just the Ca, ensures that we have not underestimated the K~{\sc xviii} triplet flux in the manner suggested by \citet{Jeltema:14}.

\begin{figure}[]
\centering
\vspace{3mm}
\hspace{-4mm}\includegraphics[width=9.5cm, angle=0]{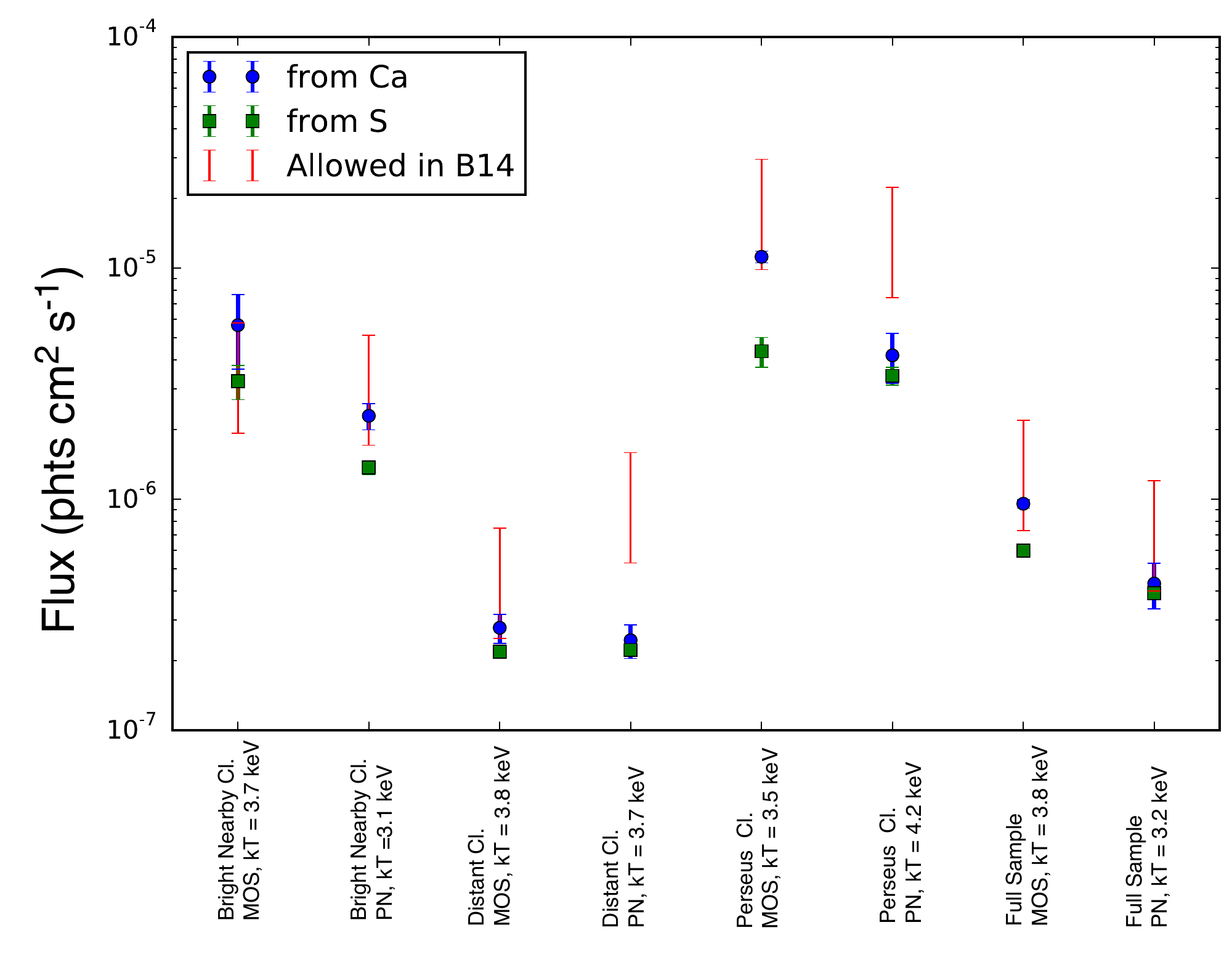}
\caption{The estimated flux in the K {\sc xviii} triplet based on single temperature plasmas from the Bu14 samples. Blue - taken from Ca emission. Green - taken from S emission. Red - the calculated (lower end) and maximum allowed (upper end) flux of the triplet in Bu14.}
\label{fig:ca-k-ratio}
\end{figure}

To demonstrate this, we have estimated the flux of the K {\sc xviii} lines using another method, the results of which are shown in Figure~\ref{fig:ca-k-ratio}. For each object listed in Bu14, the temperatures have been derived from the Ca {\sc xix} to Ca  {\sc xx} line ratio. These all lie in the range 3.1 keV to 4.2 keV. In those objects where we had extracted the S {\sc xv} flux, the temperatures from the S {\sc xv} to S {\sc xvi} ratios were also found to lie in this range.

For each object, the emissivity of the K {\sc xviii} triplet has been calculated assuming that the plasma is a single temperature component plasma at the calculated temperature, and that S, Ca and K are in collisional ionization equilibrium and they all have solar photosphere abundance (Anders \& Grevesse 1989). By comparing the predicted flux ratios with the observed flux in the Ca and S lines, we produce estimated fluxes for the K {\sc xviii} triplet based on the Ca and S observations. Error bars indicate the range of fluxes implied by the 90\% uncertainty in the Bu14 line fluxes. The red lines in the same figure show the range between the upper limit for the K {\sc xviii} flux calculated in that paper and that value with the factor of 3 safety margin included.

As can be seen, in the case of the MOS observations of the brightest clusters (Coma+Centaurus+Ophiuchus), the data shows that we have been conservative in our estimates of the maximum K {\sc xviii} flux, with estimates from this technique consistently falling at least a factor of two below the allowed values in Bu14.

\section{Comment on ``Where do the 3.5 keV photons come from? A morphological study of the Galactic Center and of Perseus"}
\label{app:carlson}
\cite{Carlson:14} presents a morphological investigation of the $\sim$3.5~keV signal in the Galactic Center (GC) and the Perseus cluster, concluding that in a template-based maximum-likelihood approach neither object prefers a dark matter-like contribution.

However, using templates that are derived directly from a few broad energy bands of the data essentially reduces the spectral information that is available. Since the $\sim$3.5~keV flux in clusters is of order 1\% of the continuum at {\it XMM-Newton}'s spectral resolution, it is essential to determine the continuum emission to better than 1\%. This is a non-trivial exercise even in a forward modeling approach as done in Bo14 and Bu14, and is impossible in the template approach. This can be seen from the broad brackets of continuum models in Figures 5 and 6 of \cite{Carlson:14}. If a continuum template is incorrect by more than a percent at 3.5 keV (which is almost a certainty), the $\sim$3.5~keV line contribution to the residual signal would be very subdominant, the residuals will be dominated by astrophysical components and, of course, follow the spatial distribution of the astrophysical templates, biasing the results against dark matter-like behaviour.

It should be noted in addition that the detection of the $\sim$3.5~keV signal in Bu14, using the same \xmm MOS data as \cite{Carlson:14}, has a significance of only about 3.4$\sigma$ for the integrated data of the entire field of view (excluding the 1$^\prime$ cluster core). Given such low significance for the whole cluster, it is difficult to see how it would be possible to subdivide the dataset and obtain statistically significant measurements of the spatial behaviour of the line signal, as is for example suggested by the size of the error bars in Figure 6 of \cite{Carlson:14} or by their discussion of the perceived `clumped nature' of the residuals in Section 3.1. The errors on the actual $\sim$3.5~keV line contribution in various sub-regions are likely understated.

Lastly, the effect of absorption by the intervening interstellar medium on the GC analysis is strongly underestimated in \cite{Carlson:14}. They use the HI data to estimate the absorbing column density, concluding that absorption at 3.5 keV is insignificant (a few percent effect). While these data are adequate over most of the sky, at low Galactic latitudes the true X-ray absorption is often higher. Indeed, using {\em Chandra}\/ X-ray spectra for the GC fields, \cite{Muno:2004bs} and \cite{Muno:2004wh} measure the absorption column densities for various diffuse emission regions and for various point sources, respectively. They find median column densities close to $6\times10^{22}$ cm$^{-2}$, while between 30 and 50\% of the analyzed area has $N_H > 10^{23}$ cm$^{-2}$. This is much higher than the HI-based value; the excess can be due to molecular gas, etc.  At 3.5~keV, such values of $N_H$ correspond to attenuation by factor 2--3, not a few percent. These X-ray absorption measurements are directly applicable here, and were used in Bo14. This impacts any upper limits computed for dark matter decay. In addition, the absorption is likely irregularly distributed over the GC area (for example, the giant molecular clouds align with the Galactic plane), making an isotropic dark matter template inadequate.

\section{Comment on ``A novel scenario for the possible X-ray line feature at $\sim$3.5~keV: Charge exchange with base sulfur ions''}
\label{app:cx}
Recently, \citet{gu:15} suggested that the unidentified $\sim$3.5~keV line could have originated from via charge exchange between bare sulfur and neutral hydrogen interacting with a relative velocity of $\sim 200$\,km/s.  New calculations of this interaction \citep{mullen:16} suggest that the dominant cross sections are to the $9p$\ and $10p$\ excited states of S XVI, leading to transitions at 3.45 keV and 3.46 keV, respectively.  Although at lower energies, we agree that if present these transitions could affect the fits to the cluster spectra, as noted by \citet{gu:15}.  However, \cite{gu:15} also argues that these S XVI transitions are a ``unique feature for probing CX in hot astrophysical plasmas,'' at least at CCD resolution.  Although possibly true in the X-ray band, CX at the level implied by \citet{gu:15} should also create detectable hydrogen H$\alpha$\ emission, although of course CX is not the only mechanism that could generate this line.  The relationship between H$\alpha$\ and X-rays in clusters has been studied extensively; for example, \citet{fabian:03} found that the H$\alpha$\ filaments in the Perseus cluster, which extend about 2 arcminutes in radius from the core, are associated with soft X-rays with a temperature of $\sim 0.9$\, keV. 

To calculate the possible H$\alpha$\ flux, we will assume a typical cluster sulfur abundance of 1/3rd solar, or [S/H] = 6.72.  Between 2-4 keV, the fractional population of fully stripped S$^{16+}$\ varies between $\sim 0.42 - 0.84$; for concreteness, we use the value at 3 keV, $0.72$, which also corresponds to the 200 km/s velocity where \citet{mullen:16} find the S$^{16+}$\ cross section peaks in the key $9p$\ and $10p$\ states.  Inherent in the assumption that CX is occurring is that somehow the cluster contains a hot plasma mixing with a cool neutral plasma, possibly due to a cool infalling filament that is slowly ``leaking'' neutral hydrogen. In this case, the ionized hydrogen and neutral hydrogen can also interact, either via excitation or CX.  To completely calculate the resulting H$\alpha$\ emission would require a complete level population calculation; we use a simpler approximation to this from \citet{mclaughlin:99},
\begin{equation}
\sigma(H\alpha) = \sigma(1s \rightarrow 3s) + 0.118\sigma(1s \rightarrow 3p) + \sigma(1s \rightarrow 3d).
\end{equation}
For the excitation and charge exchange cross sections, we use values from Table V of \citet{winter:09} at 3 keV, finding a total $\sigma(H\alpha) = 2.6\times10^{-18}$\,cm$^{-2}$.  \citet{mullen:16} (via private communication) gives $\sigma(S^{16+}+H\rightarrow S^{15+}(9p,10p) + H^+) \sim 3.3\times10^{-15}$\,cm$^{-2}$.  As the CX lines are at 3.45~keV, not 3.5~keV, they are not a one-for-one replacement for the $\sim$3.5~keV feature, but rather would impact the fits in this region in some complex fashion. It is reasonable to assume that any impact would become significant when the CX line had a similar flux as the $\sim$3.5~keV feature; in this case, we find:  
\begin{equation}
F(H\alpha) = {Ab(H)\over{Ab(S)}} \times {{\sigma(H\alpha) \times F_{3.5}}\over{\sigma(S^{16+}+H\rightarrow S^{15+}(9p,10p) + H^+)}}  
\end{equation}
\noindent or $F(H\alpha) \approx 150\times F_{3.5}.$
For Perseus, Bu14 found a range of values for $F_{3.5}$\ depending upon the analysis approach. We use here the \xmm MOS values found after excluding the core 1 arcminute radius, $2.1 (+1.1,-1.0)\times10^{-5}$\,ph cm$^{-2}$s$^{-1}$\ (90\% errors).  This implies that any potential cool plasma interaction would create H$\alpha = 3.2(+1.8,-1.7)\times10^{-3}$\,ph cm$^{-2}$s$^{-1}$.   \citet{conselice:01} mapped all of the H$\alpha$\ filaments in Perseus, finding a total flux of $3.2\times10^{-13}$\,erg cm$^{-2}$s$^{-1}$, or 0.11 photons cm$^{-2}$s$^{-1}$. However, the majority of this emission was found within 1 arcminute (21 kpc) of the core.  Excluding these points, however, reduces the observed flux to $9.5\times10^{-3}$ photons cm$^{-2}$s$^{-1}$. 

Most of the filamentary H$\alpha$\ emission in Perseus must be created by other mechanisms within the filaments (collisional excitation, recombination, or photoionization), and not CX; otherwise, the bare sulfur CX line at 3.45 keV would be orders of magnitude stronger than it is.  Similar conclusions are reached by e.g. \cite{Fabian:11} through different methods. By the same token, in the core of Perseus, CX could create both a 3.45 keV line and trace H$\alpha$\ emission that could not be detected. In other words, the H$\alpha$ emission in the core of Perseus does not exclude a CX interpretation of the $\sim$3.5~keV line in the core. However, neither does the H$\alpha$ measurements necessarily indicate that CX has to be responsible for either (part of) the 3.5~keV line or the H$\alpha$ emitted at a flux as calculated above.
Rather, we notice that the filamentary flux drops off much more rapidly from the core (more than an order of magnitude at 1 arcminute radius) than the $\sim$3.5~keV line, which only drops by a factor of 2, and suggest that this may be a distinguishing characteristic to be used in the future. More work is needed, both in the laboratory to test the theoretical CX calculations, and observational to compare the radial distributions of H$\alpha$\ emission in other clusters with the core-excluded $\sim$3.5~keV line, to conclusively identify the impact of CX.

\section{Details of the Spectral Fits of Perseus with {\it Suzaku}}
\label{app:spectra-outskirts}

\begin{figure}
\includegraphics[width=\textwidth]{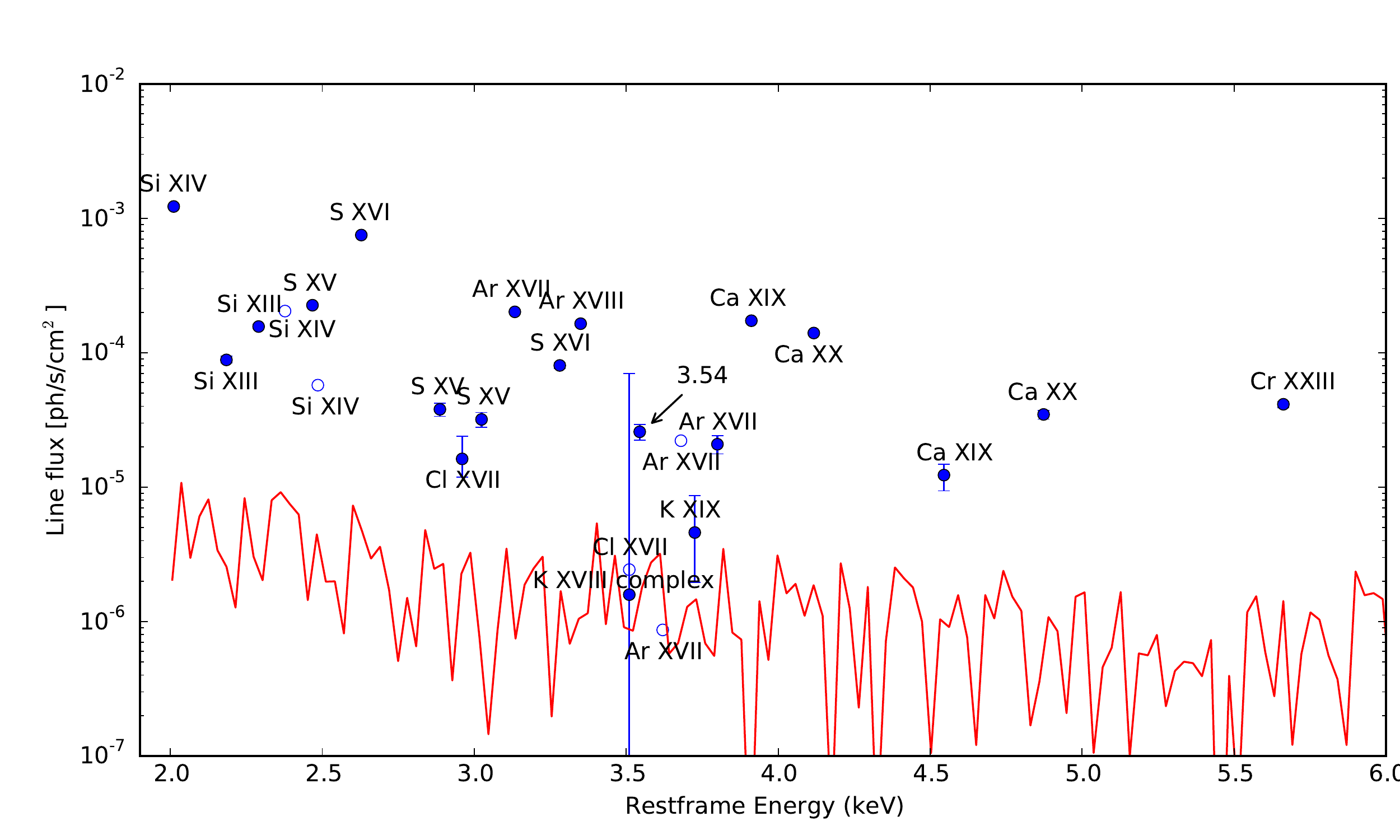}
\label{fig:atomlines-core}
\caption{The gaussian line components of the best-fit model for Region 1, in units of ph s$^{-1}$ cm$^{-2}$. Open circles indicate that the line flux was tied to another line in the fit. The red line indicates the residual level as the absolute value of the residuals in bins of 30 eV. Error bars are 1$\sigma$ obtained with the \tt error \rm command in {\sc xspec}. Note that the two lines at $\sim$3.51~keV are the Cl~{\sc xvii} line which is tied to the line of the same ion at 2.96~keV, and the K~{\sc xviii} complex whose maximum allowed flux is actually much lower than the formal error bar indicates (the maximum allowed flux is roughly 2.4 pht cm$^{-2}$ s$^{-1}$ as indicated by Tables~\ref{table:estflux} and~\ref{table:estflux_AG} and in Section~\ref{sec:reduction}). }
\end{figure}

This appendix shows the additional details and figures of the fits for all of the \suz regions used in this work. 
Table~\ref{table:bestfit-core} shows the best-fit parameters of the subregions 1a through 1c, while Figure~\ref{fig:atomlines-core} indicates graphically the best-fit gaussian line components for the best fit of Region 1.

Figures~\ref{fig:spec23} and~\ref{fig:spec424} show the spectra and residuals for all outskirt regions described in the text, being Regions 2, 3, 4, and 2-4. 

{
\begin{table*}[ht!]
\begin{center}
\caption{Same as Table~\ref{table:estflux} but for \cite{Anders:1989} solar abundances: sstimated fluxes of the Cl \textsc{xvii}, K \textsc{xviii}, Ar DR \textsc{xvii} lines are in the units of $10^{-8}$ pht cm$^{-2}$ s$^{-1}$ from AtomDB. The fluxes (and not the temperature) in this table are dependent on the assumed solar abundance, and are employed in the fits by setting the upper and lower allowed limits for the fitting procedure to 3 times and 0.1 times this flux, respectively. }
\renewcommand{\arraystretch}{1.4}
\begin{tabular}{lccccccccc}
\hline\hline
Parameter  	& Reg 1 & Reg 1a & Reg 1b & Reg 1c	& Reg 2	& Reg 3	& Reg 4 & Reg 2-4	\\
				&  \\\hline
kT based on S (keV) & 3.13 & 2.97 & 3.18 & 3.25 & 3.74 & 2.37 & 2.47 & 3.60 \\ 
kT based on Ca (keV) & 4.02 & 3.65 & 3.92 & 4.85 & 4.77 & -- & -- & 4.14 \\ 
Flux of Cl~{\sc XVII} at 2.96~keV & 1571.18 & 882.46 & 868.90 & 415.20 & 50.52 & 5.52 & 2.19 & 11.00 \\ 
Flux of Cl~{\sc XVII} at 3.51~keV & 240.05 & 133.96 & 133.00 & 63.73 & 7.88 & 0.81 & 0.32 & 1.71 \\ 
Flux of K~{\sc XVIII} at 3.47~keV & 227.78 & 138.35 & 122.66 & 56.43 & 5.32 & 1.13 & 0.43 & 1.25 \\ 
Flux of K~{\sc XVIII} at 3.49 keV & 112.44 & 68.27 & 60.59 & 27.90 & 2.65 & 0.57 & 0.22 & 0.62 \\ 
Flux of K~{\sc XVIII} at 3.51~keV & 471.09 & 280.16 & 255.33 & 118.56 & 11.78 & 2.13 & 0.82 & 2.73 \\ 
Flux of Ar DR {\sc XVII} at 3.62~keV & 66.87 & 44.72 & 35.00 & 15.46 & 1.14 & 0.58 & 0.20 & 0.29 \\ 

\hline\hline
\end{tabular}
\label{table:estflux_AG}
\end{center}
\end{table*}
}

\begin{figure}
\includegraphics[width=\textwidth]{{fD1a}.pdf}
\includegraphics[width=\textwidth]{{fD1b}.pdf}
\label{fig:spec23}
\caption{Regions 2 and 3. Showing the data and model fits to those regions, with the residuals in the bottom panel.}
\end{figure}

\begin{figure}
\includegraphics[width=\textwidth]{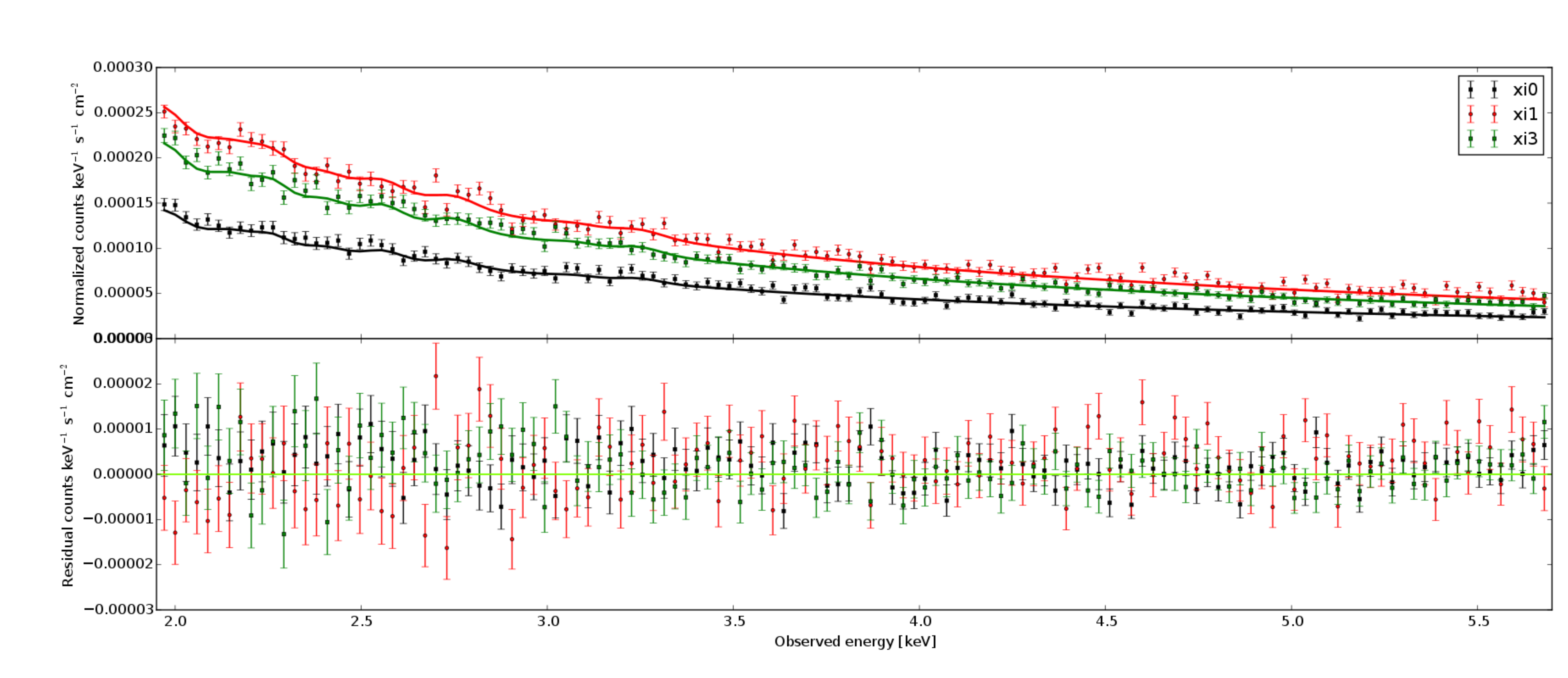}
\includegraphics[width=\textwidth]{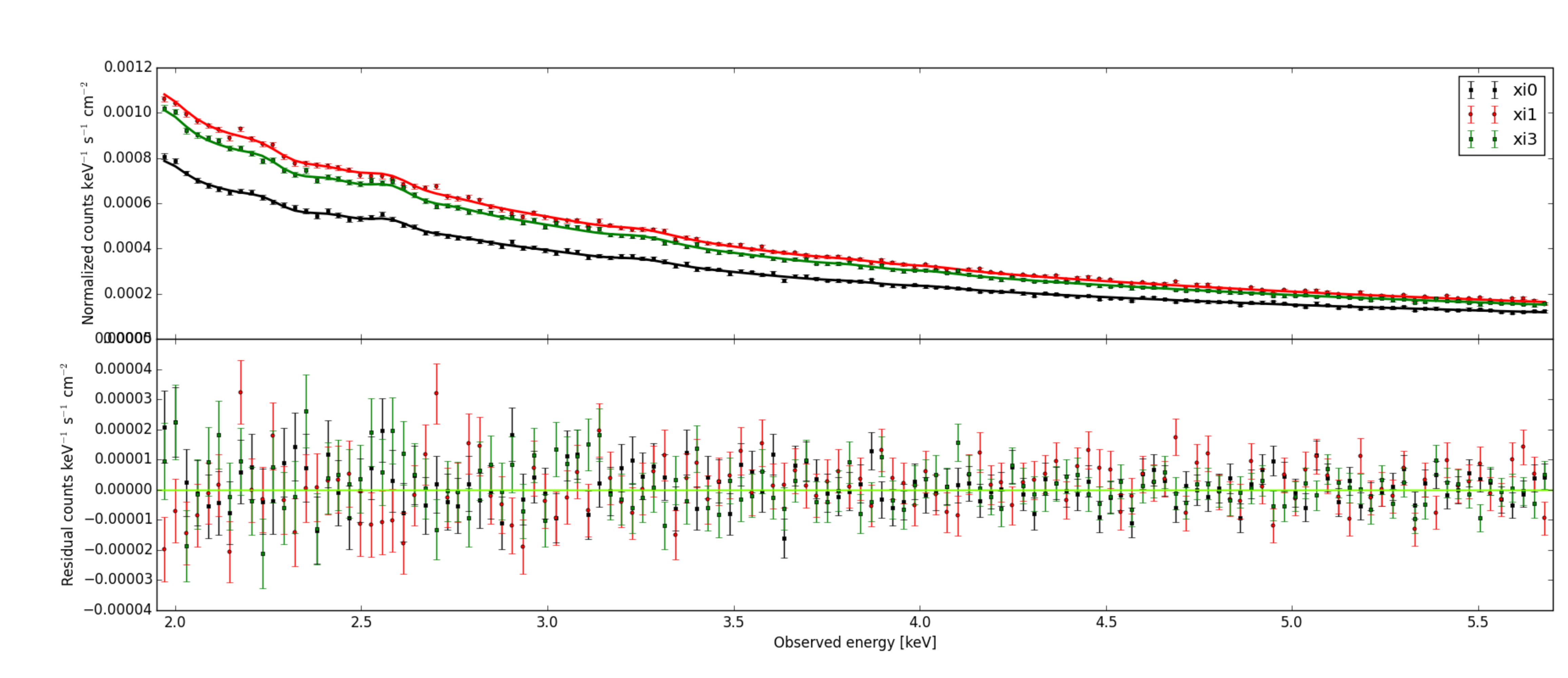}
\label{fig:spec424}
\caption{Regions 4 and 2-4. Showing the data and model fits to those regions, with the residuals in the bottom panel.}
\end{figure}

\bibliography{xrays,preamble,astro,perseus}

\end{document}